%Paper: hep-th/9401103
%From: sadov@string.harvard.edu (Vladimir Sadov)
%Date: Thu, 20 Jan 94 17:37:56 -0500
%Date (revised): Thu, 20 Jan 94 18:04:37 -0500
%Date (revised): Thu, 20 Jan 94 18:18:52 -0500
%Date (revised): Fri, 21 Jan 94 17:44:42 -0500

%
%------This TeX file needs harvmac macros-------------------------
\input harvmac
\Title{HUTP-93/A040}{Quantum cohomology of partial flag manifolds
$F_{n_1\cdots
n_k}$}
\centerline{Alexander Astashkevich}
\bigskip
\centerline{Department of Mathematics}
\centerline{M.I.T.}
\centerline{Cambridge, MA 02139}
\bigskip
\centerline{and}\bigskip
\centerline{Vladimir Sadov}
\bigskip\centerline{Lyman Laboratory of Physics}
\centerline{Harvard University}\centerline{Cambridge, MA 02138}
\centerline{and}\centerline{L.D. Landau Institute for Theoretical
Physics,
Moscow}

\vskip .3in
We compute the quantum cohomology rings of the partial flag manifolds
$F_{n_1\cdots n_k}=U(n)/(U(n_1)\times \cdots \times U(n_k))$. The
inductive
computation uses the idea of Givental and Kim \ref\Gi{A.~Givental,
B.~Kim {\it
Quantum Cohomology of Flag Manifolds and Toda Lattices} preprint
UC
Berkeley,
December 1993, hep-th 9312096.}. Also we define a notion of the
vertical
quantum cohomology ring of the algebraic bundle. For the flag bundle
$F_{n_1\cdots n_k}(E)$ associated with the vector bundle $E$ this
ring is
found.

\Date{12/93}
\newsec{Introduction and summary.}
The quantum cohomology rings are only known for some classes of
varieties.  For
  Calabi-Yau manifolds one  \ref\CaGr{P.~Candelas, P.~Green,
L.~Parkes, X.~de
la Ossa {\it A pair of Calabi-Yau manifolds as an exactly soluble
superconformal field theory} Nucl.~Phys.~ {\bf B359} (1991) 21
} can transform the problem to the ``mirror dual" one which deals
with
variations of the Hodge Structure. The latter is ``in principle"
solvable,
although the real computations in terms of Picard-Fuchs equations may
be pretty
hard. All known examples describe one- or two- parametric
deformations of the
classical cohomology rings (see for example \ref\AGM{P.~S.~Aspinwall,
B.~R.~Greene, D.~R.~Morrison {\it Calabi-Yau Moduli Space, Mirror
Manifolds and
Spacetime Topology Change in String Theory.}, preprint
IASSNS-HEP-93/38}\ref\Klemm{ Albrecht Klemm and Stefan
Theisen,{\it
Considerations of One-Modulus Calabi-Yau Compactifications:
Picard-Fuchs
Equations, K\"ahler Potentials and Mirror Maps},
TUM-TH-143-92,
KA-THEP-03/92}\ref\twom{Philip Candelas, Xenia de la
Ossa, Anamaria Font, Sheldon Katz and David R. Morrison {\it Mirror
Symmetry
for Two Parameter Models}, preprint hep-th/9308083}).

 The ring $QH^*({\bf P}^n)= {\bf C}[x]/(x^{n+1}-q)$ for the
projective spaces
has been known since long ago in physics \ref\Wicp{E.~Witten {\it
Nucl.~Phys.~}
{\bf B340} (1990) 281}  and mathematics (symplectic Floer theory
\ref\FI{A.~Floer {\it Symplectic Fixed Points and Holomorphic
Spheres.},
Comm.~Math.~Phys.~ {\bf 120} (1989) 575-611}). More generally, one
can
construct the moduli spaces of rational curves for the toric
varieties
\ref\Guest{M.~Guest {\it The Topology of the Space of Rational Curves
on a
Toric Variety}, preprint alg-geom/9301005.}. Recently Batyrev
\ref\Ba{
V.~Batyrev {\it Quantum Cohomology Rings of
Toric Manifolds},
preprint 1993,
alg-geom/9310004}
has conjectured a general formula for the quantum ring in that case.
{}From the
physical point of view Batyrev's result can be obtained using the
Hamiltonian
reduction of the linear $\sigma$-model by the real torus
\ref\WiNII{E.~Witten
{\it Phases of $N=2$ theories in two dimensions}, preprint
IASSNS-HEP-93/3}.

Another example where the hamiltonian reduction of the linear $\sigma
$-model
(by $U(n)$) does work \ref\VaNII{S.~Cecotti, C.~Vafa {\it On
Classification of
$N=2$ Sypersummetric Theories.} preprint HUTP-92/A064}, \WiNII is the
Grassmanian $Gr(n,m)=U(n)/(U(n-m)\times U(m))$, because this manifold
can be
presented  as $Gr(n,m)={\bf C}^{nm}//U(m)$. The relations in
$QH^*(Gr(n,m))$
were discussed in many physical papers \ref\gep{D. Gepner,  {\it
Fusion Rings
And Geometry,} Commun. Math. Phys.
{\bf 141} (1991) 381} \ref\geps{D. Gepner and A. Schwimmer, {\it
Symplectic
Fusion
Rings And Their Metric,} Nucl. Phys. {\bf B380} (1992)
147}\ref\gepI{ D.
Gepner,
{\it Foundations Of Rational Conformal Field Theory, I} (Cal Tech
preprint,
1992).}
\ref\vafaG{C. Vafa, {\it Topological Mirrors And Quantum Rings,}
in {\it Essays On Mirror Manifolds}, ed. S.-T. Yau (International
Press, 1992).}
\ref\intrilligator{K. Intriligator, {\it Fusion Residues,} Mod.
Phys.
Lett. {\bf A6} (1991) 3543.}\ref\WiG{E.~Witten{\it The Verlinde
Algebra And The
Cohomology Of The Grassmannian},
preprint  IASSNS-HEP-93/41,
hep-th/9312104}.
They are proven mathematically \ref\BDW{A.~Bertram, G.~Daskalopolous,
R.~Wentworth {\it Gromov Invariants for Holomorphic Maps from Riemann
Surfaces
to Grassmanians}, Harvard preprint, April 1993}  for the Grassmanians
of
2-planes.

Recently Givental and Kim \Gi have  computed  the quantum cohomology
ring of
the complete flag manifold $F_n=U(n)/ (U(1))^{\times n}$. The main
idea of
their paper was to use the functoriality properties of the {\it
equivariant}
quantum cohomology. Here we extend the arguments of \Gi to cover all
{\it
partial} flag manifolds  $F_{n_1\cdots n_k}=U(n)/(U(n_1)\times \cdots
\times
U(n_k))$ ``interpolating" between Grassmanians and complete flag
manifolds.

In general, the quantum cohomology is rather difficult to compute. An
obstacle
for developing any efficient computational technique is our lack of
understanding of its functoriality properties. In the classical
Algebraic
Topology, whenever there is a map $i:\ X\rightarrow Y$ between two
manifolds,
there is a pullback morphism $i^*$ of their cohomology rings such
that
$i^*H^*(Y)$  --- a pullback of cohomology of $Y$ --- is a subring in
$H^*(X)$.

For the quantum cohomology rings, we know no analogs of the pullback
morphism.  As an illustration, let us consider a simple example.
Take a  map $i:\ {\bf P}^1 \rightarrow {\bf P}^2$ of degree one, then
cohomology of ${\bf P}^1$ is generated by the pullback $x=i^*y$ of
the K\"ahler
class $y$ of ${\bf P}^2$; this is a particular case of functoriality
with
respect to the maps between manifolds. Now if we go to the {\it
quantum
cohomology rings}, we see there is a relation $x^2 = 1$ in $QH^*({\bf
P}^1)$
and another one  $y^3 = 1$ in $H^*({\bf P}^2)$. There are no
non-trivial
morphisms between such rings. Here, as in general, it is impossible
to define a
meaningful pullback for the quantum cohomology.

There is a natural object defined for the algebraic bundles $\pi
:X\rightarrow
B$ with fiber $F$ a K\"ahler manifold. We call it the vertical
quantum
cohomology ring $QH_V^*(X,B,F)$. Its classical limit coincides with
the
cohomology ring $H^*(X)$ of the total space. The quantum
multiplication in
$QH_V^*(X,B,F)$ is a deformation of multiplication in
$H^*(X)$ by means of the {\it vertical} rational curves $\Sigma$ in
$X$, i.~e.~
such that $\pi (\Sigma )=\{pt\}$. The ring $QH_V^*(X,B,F)$ contains
$H^*(B)$ as
a subring and can be considered as a $H^*(B)$-algebra. The vertical
cohomology
{\it is} functorial  with respect to the {\it base change} morphisms
$B
\rightarrow B'$.

A paper has the following structure:

The second section both introduces the basic notions and outlines the
main
results of the whole paper. In particular, the ring
$QH^*(F_{n_1\cdots n_k})$
is described in the Section $2.3$  The section is written using in
the physical
language and is primarily intended for a physicist reader.

The third section discusses essentially the same topics as the first
one. More
mathematically oriented reader will find there the formal definitions
and
properties of the objects used in course of subsequent computations.

In the fourth section we discuss the equivariant quantum cohomology,
first
defined in \Gi. In Sec.~$4$ we give another definition, of more
algebraic
nature.

The fifth section deals with the computation of the ring
$QH^*(F_{n_1\cdots
n_k})$. This inductive computation is based on the ideas of \Gi. The
base of
induction is provided by the known answer for the quantum cohomology
of
Grassmanians $U(n)/(U(n_1)\times U(n_2))$. The latter is discussed in
many
papers \gep---\WiG which we summarize in Sec.~$5.2$. A step of
induction uses
the functoriality properties of the equivariant quantum cohomology
discussed in
Sec.~4.

{\it Acknowledgements.} A.~A.~is very grateful to R.~Bezrukavnikov,
D.~Kaledin,
B.~Kostant, M.~Verbitsky and D.~Vogan for valuable discussions.
V.~S.~would
like to thank M.~Bershadsky  and C.~Vafa for the helpful discussions
and
support.  We also acknowledge the interesting discussions with
S.~Piunikhin.
Both of us are grateful to Michael Bershadsky for hospitality during
the period
the major part of this work was being written.

Research of V.~S.~was supported in part by the Packard Foundation and
by NSF
grant PHY-87-14654

\newsec{A physical introduction to vertical quantum cohomology.}
\subsec{Quantum cohomology}
Quantum cohomology rings are interesting from both the physical and
the
mathematical points of view. A physicist would say they compute the
topological
correlation functions in a 2-d $N=2$ supersymmetric $\sigma$-model
with a
K\"ahler manifold $X$ as a target space.

The (one-loop) beta-function of the $\sigma$-model is proportional to
the Ricci
tensor $R_{i\bar{j}}$; its cohomological class represents the first
Chern class
$c_1(X)$ of the tangent bundle $T_*X$. If $c_1(X)>0$, the
renormalization group
flow has a stable $UV$ fixed point ({\it large volume  limit})
\ref\AGG{L.~Alvarez-Gaum\'e, P.~Ginsparg {\it Comm.~Math.~Phys.} {\bf
102}
(1985) 311} where the metrics $G_{i\bar{j}}$ becomes (cohomologicaly)
the
infinite volume K\"ahler-Einstein metrics and the $\sigma$-model
becomes
(asymptotically) free; here the positivity of the first Chern class
means that
$c_1(X)$ belongs to the {\it K\"ahler cone} in $H^2(X)$.

An important class of observables in $N=2$ $\sigma$-models consists
of chiral
operators \ref\LVW{W.~Lerche, C.~Vafa, N.~Warner {\it Chiral Rings in
N=2
Superconformal Theory} Nucl.~Phys.~ {\bf B324} (1989) 427 },  which
are in
one-to-one correspondence with the de Rham cohomology classes of the
target
space $X$. These operators form a closed subtheory; the correlation
functions
in this subtheory are called topological correlation functions.
In the large volume limit the topological correlation functions are
given by
the intersection numbers in the cohomology ring $H^*(X)$. The
nonrenormalization theorems on $N=2$ supersymmetric theories imply
that away
from the $UV$ fixed point the only possible corrections to these
correlation
functions are instanton (semi-classical) corrections.

 The instantons are simply the holomorphic maps from the world sheet
into the
target space; the usual fermionic number anomaly argument shows that
in the
generic situation only {\it isolated} maps are to be counted. The
instanton sum
goes over all possible homological types of these maps. Every summand
of a
given homological type has an exponential factor $\exp{(-A)}$ where
$A$ is an
area of the worldsheet in the pullback metrics\foot{This area depends
only on
the homological type of the holomorphic map.}, and a pre-exponential
factor
proportional to the number of the isolated instantons.

The topological correlation functions can be exactly computed in
terms of the
{\it topological} $\sigma$-model \ref\Wism{E.~Witten {\it Topological
sigma
model} Comm.~Math.~Phys.~ {\bf 118}(1988) 411}\ref\Wismt{E.~Witten
{\it Mirror
manifolds and Topological Field Theory} in Essays in Mirror Symmetry,
ed.
S.~-T.~Yau, 1992}\ref\BaS{L.~Bailieu, I.~Singer {\it The topological
sigma
model.} Comm.~Math.~Phys.~ {\bf 125} (1989) 227}, related to the
original $N=2$
$\sigma$-model by what is called in \Wismt a topological $A$--twist.
The
fundamental fields of the model are
\eqn\contab{\eqalign{
{\rm Bosons:\;} & {\rm world\; sheet\; scalars\;} X^i\; - {\rm
coordinates\;
in\; the\; target\; space\; } \cr
        & {\rm world\; sheet\; 1-form\;} F_\alpha ^i - {\rm target\;
space\;
vector\;} \cr
{\rm Fermions:\;} & {\rm world\; sheet\; scalar\;} \chi ^i - {\rm
target\;
space\; vector\;} \cr
          & {\rm world\; sheet\; 1-form\;} \rho _\alpha ^i - {\rm
target\;
space\; vector.} \cr
}}
The field $F_\alpha ^i$ is what is called the auxiliary field. Both
$\rho
_\alpha ^i$ and $F_\alpha ^i$ satisfy a self-duality constraint
\eqn\sfdu{\eqalign{
&\rho ^i_\alpha =i\epsilon _\alpha ^{\ \beta}J^i_{\ j}\rho ^j_\beta
\cr
&F^i_\alpha =\epsilon _\alpha ^{\ \beta}J^i_{\ j}F^j_\beta \cr
}}
The only physical operators in the topological model are the chiral
operators
of the $N=2$ model. They represent the nontrivial cohomology classes
of the
BRST operator Q which acts on the multiplet \contab  as
follows\foot{We impose
a mass shell condition $F_\alpha ^i=0$ on the auxiliary field.}
\eqn\brst{\eqalign{
&[Q,X^i]=i\chi ^i \cr
&\{Q,\chi ^i\}=0 \cr
&\{Q,\rho ^i_\alpha \}=\partial _\alpha X^i+\epsilon ^{\
\beta}_\alpha  J^i_{\
j}\partial _\beta X^j -i\Gamma ^i_{jk}\chi ^j\rho ^k_\alpha \cr
}}
 The physical operators are in one-to-one correspondence with the de
Rham
cohomology classes of $X$: given  that a differential form $\omega
=\omega
_{i_1\cdots i_k}(X)dX^{i_1}\ldots dX^{i_k}$ represents a class
$[\omega ]\in
H^k(X)$ of de Rham cohomology, an operator
\eqn\oper{
{\cal O}_\omega = \omega _{i_1\cdots i_k}(X(z,\bar{z}))\chi
^{i_1}(z,\bar{z})\ldots \chi ^{i_k}(z,\bar{z})
}
represents a class of BRST cohomology.

Often it is more convenient to consider the operator product algebra
of the
chiral states, a {\it chiral ring} or a {\it quantum cohomology
ring}, instead
of the correlation functions. If $\{x_i\}$ is a basis of chiral
states, then
the ring structure can be written in terms of the three-point
correlation
functions on sphere as follows:
\eqn\alg{
a\cdot b =\langle a\,b\,x^i\rangle x_i
}
where $\{x^i\}$ is a dual basis with respect to the two-point
functions:
\eqn\twp{
\langle x_i\,x^j\rangle =\delta _i^j
}
The associativity of the multiplication \alg  follows from the {\it
duality}\foot{This duality has nothing to do with \twp} of the
four-point
correlation functions. For a more profound review of the physics of
the
$\sigma$-models on the K\"ahler manifolds we refer our reader to
\Wismt and
references therein.

\subsec{Vertical quantum cohomology.}
In this paper  we discuss the functoriality properties of what we
call the
vertical quantum cohomology. This object will be rigorously defined
in the
subsequent section; here we introduce it in one particular setup
using the
language of the topological $\sigma$-model. Loosely speaking, the
vertical
quantum cohomology appear in a situation when there is an algebraic
bundle or
more generally, a family of deformations with a ``quantum" fiber and
a
``classical" base. The classical  limit (large radius  limit for the
fiber) of
the vertical quantum cohomology ring is just the ordinary de Rham
cohomology
ring of the total space.

Let $X$ be a simply connected compact K\"ahler manifold with a
structure of a
locally trivial algebraic bundle over a simply connected compact
K\"ahler
manifolds $B$ (the base) with fiber $F$ and  with a projection of the
total
space $\pi :X\rightarrow B$.
In this setup, according to Deligne \ref\Del{P.~Deligne {\it
Th\'eor\`eme de
Lefschetz et crit\'eres de d\'eg\'en\'erescence de suite spectrales},
Publ.~Math.~  IHES, Vol 35(1968), pp. 107-126}, \ref\GH{P.Griffits,
J.Harris,
{\it Principles of algebraic geometry.}
Wiley, N.Y., 1978.}, the Leray spectral sequence for the bundle $\pi
:X\rightarrow B$ degenerates in the second term. This means that
$H^*(B)$ is a
subring in the cohomology ring $H^*(X)$ of the total space  and that
as a
$H^*(B)$-algebra $H^*(X)$ is generated by some elements $x_1,\ldots
,x_N\in
H^*(X)$ such that when restricted to a fiber they generate the whole
ring
$H^*(F)$.

We assume also that the  first Chern classes $c_1(F)$ and $c_1(X)$
are
non-negative. The result of multiplication of two elements in the
quantum
cohomology ring $QH^*(X)$ is represented by a power series in {\it
quantum
deformation parameters} $q_1,\ldots ,q_{n+m}$.  From the point of
view of the
topological $\sigma$-model a parameter $q_i$, corresponding to a
certain
homological class $[\Sigma ]_i$ of algebraic curves in $X$, is an
exponential
function
\eqn\defo{
q_i=\exp{(-\int_{\Sigma}k)}}
of the minus area of the curve in that class.
A semi-group\foot{A semi-group $H_{11}(X)$ is generated by the
homological
classes of algebraic curves in $X$ taken with nonnegative integer
coefficients.
To a group operation in $H_{11}(X)$ there corresponds a
multiplication in the
semi-group ring.} ring ${\bf C}[H_{11}(X)]$ is canonically imbedded
into the
ring of power series generated by the deformation parameters ${\bf
K}={\bf
C}[[q_1,\ldots ,q_{n+m}]]$. It is convenient to think of the ring
$QH^*(X)$ as
a ${\bf K}$-algebra.

Now we wish to utilize a structure of $X$ as a bundle over $B$ in a
simple way:
we take a ``partial classical limit" of the ring $QH^*(X)$ ``along
the base".
To define this limit let us take a form $k_h\in H^{11}(B)\subset
H^{11}(X)$ in
the interior of the K\"ahler cone of $X$ and let us consider a
one-parametric
family of the K\"ahler classes represented by $k(\lambda )=k+\lambda
k_h$,
$\lambda \geq 0$. Then the ``partial classical limit" corresponds to
\eqn\lavl{
\lambda \longrightarrow \infty
}
A semi-group $H_{11}(X)$ has a sub-semi-group ${\cal V}$
generated\foot{In fact
${\cal V}$=$H_{11}(F)$.} by the homological classes of curves taking
zero
values on $k_h$. These are {\it vertical curves}: the projection $\pi
:X\rightarrow B$ send them to points on the base.
There is an ideal $I$ in the semi-group ring of $H_{11}(X)$ generated
by the
set-theoretical complement to ${\cal V}$ in $H_{11}(X)$:
\eqn\ide{
I={\bf C}[H_{11}(X)\setminus {\cal V}]
}
It consists exactly of those polynomials in $q_1,\ldots ,q_{n+m}$
that tend to
zero in the limit \lavl  if we substitute the expressions \defo  for
$\{q_i\}$.

We can define a quotient ring
\eqn\cfc{
\tilde{\bf K}={\bf C}[[q_1,\ldots ,q_{n+m}]]/I={\bf
C}[[q_{n+1},\ldots
,q_{n+m}]]
}
which is related to the semi-group ring ${\bf C}[H_{11}(F)]$ as ${\bf
K}$ is
related to ${\bf C}[H_{11}(X)]$.
The quotient
\eqn\relat{
QH_V^*(X,B,F)=QH^*(X)/I
}
is a $\tilde{\bf K}$-algebra. This is what we mean by the limit \lavl
of the
ring
$QH^*(X)$. We call a ring $QH_V^*(X,B,F)$ the {\it vertical quantum
cohomology}
ring of the bundle $\pi :X\rightarrow B$.

To investigate the limit \lavl  in more detail we can consider a
topological
(type A) model with a world sheet $\Sigma$ and a target space $X$.
The
$m$-point correlation functions there are given by the path integral
\eqn\FIT{
\langle {\cal O}_{\omega _1}\cdots {\cal O}_{\omega _m}\rangle =\int
{\cal
D}X^i{\cal D}\chi ^i{\cal D}\rho ^i_\alpha {\cal D}F ^i_\alpha
\;{\cal
O}_{\omega _1}\cdots {\cal O}_{\omega _m} e^{A[X,\chi ,\rho ,F]}
}
where the operators ${\cal O}_{\omega _i}$ are given by \oper. The
action
functional $A$ is a sum of two pieces:
\eqn\sact{
A=\int \phi ^*k+ t\bigl\{Q,\int{1 \over 2}g^{\alpha \beta}G_{ij}\rho
^i_\alpha
(\partial _\alpha X^j-{1\over 2}F^j_\beta )\bigr\}
}
The first term depends only on the homological type of the map
$\phi:\Sigma
\rightarrow X$ and on the cohomology class of the K\"ahler form $k$
of $X$. The
second term does not contribute to the correlation functions, since
it is
BRST-trivial. Therefore, it suffices to compute the integral \FIT  in
the limit
$t\longrightarrow \infty$. In this limit the measure $[{\cal
D}X^i{\cal D}\chi
^i{\cal D}\rho ^i_\alpha {\cal D}F ^i_\alpha] \;e^{A[X,\chi ,\rho
,F]}$ is
localized to the {\it holomorphic} maps\foot{Note then that the first
term in
\sact   computes the area of the world sheet in the pullback metrics
$\phi
^*G_{ij}$.}  $\phi:\Sigma \rightarrow X$.

 The correlation function \FIT  depends on $\lambda$ via the
exponential term
$\exp{(-\lambda \int_\Sigma \phi ^*k_h)}$. In the limit $\lambda
\longrightarrow \infty$
the contributions to \FIT  from all except {\it the vertical} maps
$\phi
_V:\Sigma \rightarrow X$ are suppressed by the factor
$~e^{-\lambda}$. Here the
vertical maps are defined by the homological property to annihilate
(any)
horizontal K\"ahler form. (This is equivalent to a condition that the
composition $\pi \circ \phi _V:\Sigma \rightarrow B$ maps the whole
curve
$\Sigma$ to a point: ${\rm Im}(\pi \circ \phi _V)=\{\cdot\}$).

We see that in the limit \lavl  the integration in \FIT  reduces to
the
integration over zero modes of the ``horizontal" fields plus
integration over
the vertical holomorphic maps:
\eqn\OIT{\eqalign{
&\langle {\cal O}_{\omega ^{(1)}}\cdots {\cal O}_{\omega ^{(m)}}
\rangle = \cr
&\int_{[B]} \int {\cal D}X_x^I{\cal D}\chi ^I{\cal D}\rho ^I_\alpha
\;
e^{A[X,\chi ,\rho ]}\;
\omega ^{(1)}_{i_1\cdots i_{k_1}I_1\cdots I_{l_1}}(x,
X_x(z))dx^{i_1}\ldots
dx^{i_{k_1}}\chi ^{I_1}(z,\bar{z})\ldots \chi ^{I_{l_1}}(z,\bar{z})
\cr
&\cdots \omega ^{(m)}_{j_1\cdots j_{k_m}J_1\cdots J_{l_m}}(x,
X_x(z))dx^{j_1}\ldots dx^{j_{k_m}}\chi ^{J_1}(z,\bar{z})\ldots \chi
^{J_{l_m}}(z,\bar{z})\cr
}
}
In this formula the coordinates along the base are denoted by $x^i$
and the
coordinates along the fiber by $X_x^I$. The latter are only defined
over each
single point of the base which we stressed by putting a subscript
$x$.

The space of the vertical holomorphic maps is fibered over $B$ with a
space of
holomorphic maps ${\cal M}_F=\{\Sigma \rightarrow F\}$ as a fiber.
The second
integration in \OIT  is performed with $x$ fixed; it goes over the
fiber ${\cal
M}_{\pi ^{-1}(x)}$. Thus its result  is a correlation function in the
topological $\sigma$-model with the fiber $F$ as a target space.

The integral $\int_{[B]}$ is an ordinary integral over the base of
the
differential form on $B$ which we call {\it the vertical correlation
function}
and denote by $\langle {\cal O}_{\omega ^{(1)}}\cdots {\cal
O}_{\omega
^{(m)}}\rangle _{(v)}$, so finally
\eqn\corr{
\langle {\cal O}_{\omega ^{(1)}}\cdots {\cal O}_{\omega
^{(m)}}\rangle
=\int_{[B]}\langle {\cal O}_{\omega ^{(1)}}\cdots {\cal O}_{\omega
^{(m)}}\rangle _{(v)}
}

The $m$-point vertical correlation functions are defined for all
$m$-tuples
${\cal O}_{\omega ^{(1)}}\cdots {\cal O}_{\omega ^{(m)}}$ of
operators in the
limit $\lambda \longrightarrow \infty$ of the topological
$\sigma$-model by the
 second integral in \OIT. From the definition it is clear that they
satisfy
\eqn\distri{
\langle \cdots{\cal O}_{\omega}\rangle _{(v)}=\langle \cdots \rangle
_{(v)}\,\wedge \omega
}
for any $\omega \in H^*(B) \subset H^*(X)$. This means that the
operator
algebra of the topological $\sigma$-model in the limit \lavl  has a
subalgebra
isomorphic to the classical cohomology ring of the base $H^*(B)$.
{\it It shows
that the vertical quantum cohomology ring} $QH_V^*(X,B,F)$ {\it is
actually an}
$H^*(B)$-{\it algebra}.

What makes the ring $QH_V^*(X,B,F)$ a rather interesting object is
that it
behaves nicely with respect to the change of the base operation.
Let $X'$ be a fiber product of $B'$ and $F$ induced from $X$ by a
morphism $i:\
B'\rightarrow B$ of the bases\foot{In other words, $X'$ is defined as
$X\times
_B B'$.}. Then there are pullback morphisms $i^*:\,H^*(B)\rightarrow
H^*(B')$
and $j^*:\,H^*(X)\rightarrow H^*(X')$. By definition, $QH_V^*(X,B,F)$
is an
algebra over $H^*(B)$ and $QH_V^*(X',B',F)$ is an algebra over
$H^*(B')$. A
pullback $j^*$ induces a pullback morphism
$j_q^*:\,QH_V^*(X,B,F)\rightarrow
QH_V^*(X',B',F)$ of vector spaces. We claim that in fact $j_q^*$ {\it
is} a
morphism {\it of rings.}

{\bf Base change for the vertical quantum cohomology}.
\smallskip \noindent
{\it
As an $H^*(B')$-algebra,~ $j_q^*(QH_V^*(X,B,F))\otimes
_{H^*(B)}H^*(B')$ is a
subalgebra in $QH_V^*(X',B',F)$.}
\medskip
{\bf Example.} The following (trivial) example demonstrates one
standard
application of the base change property. Consider an imbedding
$\{pt\}\rightarrow B$ inducing a trivial bundle $(F,\{pt\},F)$ with
vertical
quantum cohomology $QH^*_V(F,\{pt\},F)=QH^*(F)$ given by the usual
quantum
cohomology of the fiber $F$. The base change property gives
$QH^*_V(F,\{pt\},F)=QH^*_V(X,B,F)/L$ where $L$ is an ideal in
$H^*(B)\subset
QH^*_V(X,B,F)$ generated by $\oplus_{i>0}H^i(B)$. Thus we have the
identity
\eqn\sily{
QH^*(F)=QH^*_V(X,B,F)/L
}
which relates a partial classical limit of quantum cohomology of the
total
space $X$ and the quantum cohomology of the fiber $F$.

{\bf Remark.}
Actually for the definition of the vertical quantum cohomology it is
not
necessary that the base $B$ be K\"ahler or compact. For any algebraic
$B$, it
is possible to define this ring and it still has a structure of
$H^*(B)$-algebra.
For a physicist, the most natural example would be a family of
Calabi-Yau
manifolds considered over some open set in the moduli space of their
complex
structures. But in this paper we discuss another example based on
geometry of
the algebraic vector bundles.

\subsec{Equivariant quantum cohomology and quantum cohomology of the
partial
flag manifolds.}

An interesting application of the base change property appears in a
situation
when $X$ is a bundle of partial flag manifolds $F_{n_1\cdots n_k}(E)$
associated with a complex holomorphic rank $n$ vector bundle $E$ on
$B$; the
partial flag manifold is defined as the quotient $F_{n_1\cdots
n_k}={U(n)/(\otimes_{i=1}^k U(n_i))}$. In particular, this may be a
projectivization ${\bf P}(E)$. There is a classical theory of
cohomology of
such bundles \ref\BT{R.~Bott, F.~Tu {\it Differential Forms in
Algebraic
Topology}} which effectively reduces to the following statement.
Consider a map
$i: B\rightarrow BU(n)$ of the base $B$ of $E$ to the classifying
space
$BU(n)$; then the ring $H^*(F_{n_1\cdots n_k}(E))$ as a
$H^*(B)$-algebra is
obtained as a pullback $i^*$ of the cohomology ring of the universal
$F_{n_1\cdots n_k}$ flag bundle on $BU(n)$. This latter ring is what
is called
the {\it $U(n)$-equivariant cohomology} $H^*_{U(n)}(F_{n_1\cdots
n_k})$ of the
flag variety. The ring $H^*_{U(n)}(F_{n_1\cdots n_k})$ is an algebra
over
$H^*(BU(n))={\bf C}[c_1,\ldots ,c_n]$ which is a graded polynomial
algebra
generated by the Chern classes $c_1,\ldots ,c_n$ of the vector
bundle, associated to the universal bundle
$EU(n)\rightarrow BU(n)$. The pullback morphism $i^*:
H^*(BU(n))\rightarrow
H^*(B)$ maps the generators $c_i\longrightarrow c_i(E)$ to the Chern
classes of
the bundle $E$.

For example, for the projective space ${\bf P}^{n-1}$ the equivariant
cohomology ring is given by
\eqn\cpvs{
H^*_{U(n)}({\bf P}^{n-1})=H^*(BU(n))[x]/(x^{n}+c_1x^{n-1} + \cdots
+c_{n})
}
Taking the pullback $i^*$, we get the cohomology ring of the
projectivization
${\bf P}(E)$:
\eqn\cpvb{
H^*({\bf P}(E))=H^*(B)[x]/(x^{n}+c_1(E)x^{n-1} + \cdots +c_{n}(E))
}
where $c_i(E)$ is the i-th Chern class of $E$.

The base change property of the vertical quantum cohomology ensures
the same
holds true in quantum case, if one uses {\it quantum equivariant}
cohomology
instead of equivariant ones. It is the equivariant cohomology of the
complete
flag manifolds $F_n$ that was computed in \Gi. We can formulate the
definition
of \Gi in terms of vertical quantum cohomology as follows: for any
$G$-manifold
$F$ the quantum equivariant cohomology ring is defined as \eqn\qec{
QH_G^*(F)=QH^*_V(F\times _G EG, BG,F)
}

The ring $QH^*_{U(n)}(F_{n_1\cdots n_k})$ will be computed in section
5.  To
describe it, let us introduce the generators $x_1^{(1)},\ldots
,x_{n_1}^{(1)},\ldots ,x_{1}^{(k)},\ldots ,x_{n_k}^{(k)}$ of degrees
${\rm
deg}\ x_i^{(j)}=2i$ and the quantum deformation parameters
$q_1,\ldots,
q_{k-1}$ of degrees ${\rm deg}\ q_i=n_i+n_{i+1}$. The ring
$QH^*_{U(n)}(F_{n_1\cdots n_k})$ is a quotient  by a homogenous ideal
$QI$ of
the  graded polynomial algebra generated over $H^*(BU(n))$ by
$\{x_i^{(j)},
q_j\}$:
\eqn\QEC{
QH^*_{U(n)}(F_{n_1\cdots n_k})=H^*(BU(n))[x,q]/QI
}
 The ideal $QI$ is generated by the coefficients of the polynomial
\eqn\reeqc{
{\rm det}[\lambda \,{\bf 1}+A]-\lambda ^{n}-c_1\lambda ^{n-1}-\ldots
-c_{n}
}
where the $n\times n$ matrix $A$ is defined as
$$\pmatrix{  &x_1^{(1)}  &\ldots  &x_{n_1}^{(1)} &0  &\ldots
&{\scriptstyle
-(-1)^{n_2}}q_1 &\ldots &0 &\ldots &\ldots &0 &0  \cr
             \noalign{\smallskip}
             &-1\hfill  &\ldots  &0 &0 &\ldots &0 &\ldots &0 &\ldots
&\ldots
&0 &0 \cr
             \noalign{\smallskip}
	     &\vdots  &\ddots  &\vdots &\vdots &\ddots &\vdots
&\ddots &\vdots &\ddots
&\ddots  &\vdots &\vdots \cr
             \noalign{\smallskip}
             &0  &\ldots  &-1\hfill &x_1^{(2)}  &\ldots
&x_{n_2}^{(2)} &\ldots
&{\scriptstyle -(-1)^{n_3}}q_2 &\ldots  &\ldots &0 &0 \cr
             \noalign{\smallskip}
             &\vdots  &\ddots  &\vdots &\vdots &\ddots &\vdots
&\ddots &
&\ddots &\ddots  &\vdots &\vdots \cr
             &\vdots  &\ddots  &\vdots &\vdots &\ddots &\vdots
&\ddots &
&\ddots &\ddots  &\vdots &\vdots \cr
             &\vdots  &\ddots  &\vdots &\vdots &\ddots &\vdots
&\ddots  &\ddots
& &\ddots  &\vdots &\vdots \cr
             &\vdots  &\ddots  &\vdots &\vdots &\ddots &\vdots
&\ddots  &\ddots
& &\ddots &\vdots &\vdots \cr
             \noalign{\smallskip}
             &0  &\ldots  &0 &0 &\ldots &0 &\ldots  &\ldots
&x_1^{(k)} &\ldots
&x_{n_k-1}^{(k)} &x_{n_k}^{(k)}   \cr
            \noalign{\smallskip}
             &0 &\ldots  &0 &0 &\ldots &0 &\ldots  &\ldots
&-1 &\ldots &0 &0   \cr
	     &\vdots  &\ddots  &\vdots &\vdots &\ddots &\vdots
&\ddots &\ddots &\vdots
 &\ddots &\vdots &\vdots \cr
	     &0  &\ldots  &0 &0 &\ldots &0 &\ldots &\ldots  &0
&\ldots &\,\,-1\hfill
&0 \cr
}$$

The entries of $A$ are graded according to the number of the diagonal
they
belong to:
 those on the diagonal just below the main diagonal have degree $0$,
those on
the main diagonal has degree $2$ etc. All the entries on the diagonal
just
below the main one equal $(-1)$ and yet below everything equal 0.
There are
non-zero entries on the main diagonal and above in the 1-st,
$n_1$-th,
$n_1+n_2$-th, $\ldots$, $n_1+\cdots +n_{k-1}$-th rows:
\eqn\matrentr{
A_{(n_1+\cdots +n_{i-1},\ n_1+\cdots +n_{i-1}+j)}=
\cases{-1,      &if $j=-1$; \cr
       x^{(i)}_j, &if $0\leq j\leq n_i$; \cr
       -(-1)^{n_{i+1}}q_i,     &if $j=n_i+n_{i+1}$; \cr
       0,       &otherwise. \cr
       }
}

Taking the pullback $i^*$, we obtain the {\it vertical} quantum
cohomology of
the bundle $F_{n_1\cdots n_k}(E)$ as an algebra generated over
$H^*(B)$ by the
same generators $x_1^{(1)},\ldots ,x_{n_k}^{(k)}$  and  quantum
deformation
parameters $q_1,\ldots, q_{k-1}$ as above with relations given by the
coefficients of polynomial
\eqn\rerel{
{\rm det}[\lambda \,{\bf 1}+A]-\lambda ^{n}-c_1(E)\lambda
^{n-1}-\ldots
-c_{n}(E)
}
Setting in \rerel  $c_1(E)=0,\ldots ,c_n(E)=0$ (cf. \sily ), one gets
the
relations in the quantum cohomology ring $QH^*(F_{n_1\cdots n_k})$ of
the
partial flag manifold $F_{n_1\cdots n_k}$.

{\bf Remark} It should be noted that the situation with the ring
$QH^*(F_{n_1\cdots n_k}(E))$ is much less clear. We discuss this in
Appendix A.

\newsec {A math introduction to vertical quantum cohomology.}
\subsec{Quantum cohomology.}
Let $X$ be a  smooth complex  projective variety, $c$  the first
Chern
class
of the tangent bundle of $X$. Now and for all we assume that $c$ is
non-negative. Let us choose some basis
$\omega_1,\omega_2,...,\omega_k$
in $H^2(X,{\bf Z})\cap{H^{1,1}(X,{\bf C})}$
such that the homology class of a complex curve $S$ is given by
a string
$d=(d_1,...,d_k)$ of its coordinates with respect to
$\omega_1,\omega_2,...,\omega_k$, with $d_1,...,d_k{\geq}0$.
Let $L$ be a sublattice in $H_2(X,{\bf Z})\cap H_2(X,{\bf C})$
 generated by the homology classes of
complex curves
(${\bf CP}^1$).
We use the notation $q^d$ for the elements
 of the group ring of the lattice $L$. Since we
have
already chosen some basis in $H^2(X,{\bf Z})\cap{H^{1,1}(X,{\bf
C})}$,
we can identify the element $q^d$ with the monomial
$q_1^{d_1}...q_k^{d_k}$.
We denote by ${\bf K}$ the ring ${\bf C}[[q_1,...,q_k]]$ (~but see
Appendix
C~).

 Let $d$ be an element of $L$. We denote by ${\cal M}_d$ the moduli
space of
degree $d$ algebraic maps ${\bf CP}^1\rightarrow {X}$ with fixed
three points
$(0,1,\infty).$  In our case
${\cal M}_d$  is a well-defined quasiprojective algebraic variety
(~see
Appendix D~). The Riemann-Roch formula
tells us
that ${ {dim}}\,{\cal M}_d{\geq}c(d)+ {\rm dim}\,X$. We assume in
this paper
that for $X$ the
equality
holds, i.e. $  {\rm dim}\,{\cal M}_d=c(d)+ {\rm dim }\,X$.

 Let $p_1,p_2,...,p_s$ be
closed differential forms on $X$ of degree $r_1,...,r_s$
respectively.
We want to define the correlation function
${\langle}p_1|p_2|...|p_s{\rangle}$
 now. For that purpose let us choose $s$ (different) generic points
$x_1,...,x_s$ on ${\bf CP}^1$. For each $d{\in}L$, we have a map
$\psi:{\cal M}_d{\rightarrow}X^s$. It is given by
$$\psi:{\phi}\,{\mapsto}
(\phi(x_1),\phi(x_2),...,\phi(x_s)){\in}X^s~~~~~~~~~where~~~~
\phi{\in}{{\cal M}_d}.$$
Note that this is an algebraic map.
We have $s$ canonical projections $\pi_i$ ( $i=1,...,s$ ) from $X^s$
to
$X$.
So we have the  differential form
$P={\pi_1}^*{p_1}\wedge{\pi_2}^*{p_2}\wedge...\wedge{\pi_s}^*{p_s}$
on $X^s$.

\vskip 3mm

{\bf Definition.}{\it
 The correlation function ${\langle}p_1|p_2|...|p_s{\rangle}$ is
defined by
\eqn\qucf{
{\langle}p_1|p_2|...|p_s{\rangle}=\sum_{d}^{} q^d \int_{{\cal
X}_d}^{}
{{\psi}^*P}.
}
}

First of all, we must show that this definition is correct. In order
to do that
we must show that the integral $\int_{{\cal M}_d}^{} {{\psi}^*P}$ is
well-defined.
But this is easy to do. Indeed, we are integrating a differential
form of
the
highest
degree, so we can integrate it over any affine subvariety $U$ of
${\cal M}_d$
such that $  {\rm dim}~({\cal M}_d-U)<  {\rm dim}~{\cal M}_d$.
Second, let us distinguish  two cases:
\item{1)} the dimension of the closure  of $\psi({\cal M}_d)$ is less
then the
dimension of ${\cal M}_d$
\item{2)} the dimension of the closure of $\psi({\cal M}_d)$ is the
same as the
dimension of ${\cal M}_d$.

In the first case it is quite obvious that the integral is equal to
0.
In the second case we can find such affine subvariety $U$ of ${\cal
M}_d$
satisfying the above condition that the restriction of $\psi$ to $U$
gives a
covering
of the image $\psi(U)$. So we have
$$  \int_{{\cal M}_d}^{} {{\psi}^*P}=\int_{U}^{} {{\psi}^*P}=
{  {deg}}~(\psi){\int_{\psi(U)}^{} {P}}.$$
The last integral is well-defined (since $\psi(U)$ is a constructible
set).

Second, we must show that the definition \qucf does not depend on the
choice\foot{We can integrate over the set of all possible choices of
points
(i.e. over ${\bf CP}^s$), so the definition does not depend on the
choice
  of points }
of $x_1,...,x_s$ on ${\bf CP}^1$. A  reader may find the proof in
\ref\Tian{G.~Tian, S.-S.~Roan, {\it A mathematical theory of quantum
cohomology}, in preparation}.

\vskip 2mm

{\bf Properties of correlation functions:}

\item{1.} They are multi-linear and skew-symmetric.
\item{2.}
${\langle}p_1|p_2|...|p_s|1{\rangle}={\langle}p_1|p_2|...|p_s{\rangle}
$
 (where 1 is considered as a zero degree differential form).
\item{3.} ${\langle}p_1|p_2|...|p_s{\rangle}|_{q=0}=(p_1,...,p_s)$
where
  $(p_1,...,p_s)=\int_{X}^{} {p_1{\wedge}p_2{\wedge}...{\wedge}p_s}$.
\item{4.} ${\langle}p{\rangle}=(p)=\int_{X}^{} {p}.$
\item{5.} $ {\langle}p_1|p_2{\rangle}=(p_1,p_2)$ is the
usual Poincar\'e pairing.
\vskip 5mm

All properties follow obviously from the definition except the last
one.
Let us explain why the last property is true.
We choose two points $x_1$ and $x_2$ of ${\bf CP}^1$. The group ${\bf
C}^*$
is the group of automorphisms of ${\bf CP}^1$ with two fixed points
$x_1$ and $x_2$ . It acts on the variety ${\cal M}_d$, and the map
$\psi$ from
the
definition is constant on the ${\bf C}^*$-orbits. The unique
non-empty
  variety ${\cal M}_d$ on which ${\bf C}^*$ acts trivially is
 ${\cal M}_0=X$. On the other moduli spaces the action is "free
almost
everywhere".
 So the dimension
of the closure of the image of $\psi$ is less then that of
${\cal M}_d$ if $d{\neq}0$. So we get the ordinary Poincar\'e
pairing.

Now we define quantum cohomology. The Poincar\'e pairing gives
us an isomorphism of $H^*(X,{\bf C})\otimes{{\bf K}}$ with its dual
as ${\bf K}$-modules. So the triple pairing gives a product:
\eqn\qupro{
(H^*(X,{\bf C})\otimes{{\bf K}})\bigotimes_{{\bf K}}(H^*(X,{\bf
C})\otimes{{\bf
K}})\rightarrow{H^*(X,{\bf C})\otimes{{\bf K}}}.
}
It is obvious that the product \qupro is skew-commutative.
We are not going to prove that this product is associative
(as it is highly non-trivial. See \Tian), but we assume that it is
true. The
product \qupro gives us what we call {\it quantum cohomology ring}
$QH^*(X,{\bf
K})$ of $X$. As ${\bf K}$-module, the quantum cohomology is
isomorphic to
$H^*(X,{\bf C})\otimes{{\bf K}}$, but
the products in these two rings are different.

\vskip 2mm

{\bf Properties \foot { see \Gi } :}

\item{1)} the product \qupro is   a $q$-deformation of the classical
product
\item{2)} the product \qupro respects the usual grading in the
cohomology
$H^*(X,{\bf C})\otimes{{\bf K}}$ (remember that ${\bf K}$ is a graded
ring).
\vskip 5mm

\subsec{ Vertical quantum cohomology.}

Let us suppose\foot{In fact, vertical quantum cohomology can be
defined in more
general situation.
But we need it only in this case and it is easier to describe it
(really the
ring ${\bf K}$) in this
particular situation.} that  we have a locally trivial algebraic
fiber bundle
$X$ with a base $B$ and a fiber $F$.
We assume that all $X,B$ and $F$ satisfy all the conditions imposed
on the
varieties
for which we have defined the quantum cohomology.
Deligne's theorem (see \Del) tells that the corresponding Leray
spectral
 sequence degenerates in $E_2$. So the map  from $H^2(X,{\bf C})$ to
$H^2(F,{\bf C})$ is surjective.
 Denote the projection from $X$ to $B$ by $\pi$ and
an inclusion of $F$ into $X$ by $i$.

Let ${\tilde{\omega}}_1,...,{\tilde{\omega}}_k$ be our chosen
 basis in $H^2(F,{\bf Z})\cap{H^{1,1}(F,{\bf C})}$ for $F$,
${\tilde{\omega}}_{k+1},...,{\tilde{\omega}}_l$ basis for $B$. We can
find a
basis
$\omega_1,...,\omega_k,\omega_{k+1},...,\omega_l$ for $X$ such that
$\omega_i=\pi^*({\tilde{\omega}}_i)$ for $i>k$ and
${\tilde{\omega}}_i=i^*(\omega_i)$ for $i{\leq}k+1$.
Certainly the choice is not unique but nothing essential would depend
on it.
We want to deform the usual multiplication using only the maps from
${\bf CP}^1$ to $X$ which project onto points  in $B$.
We call such maps $vertical$.
The homology class of such a map can be parametrized by a string
$d=(d_1,...,d_k)$, the  evaluation of $\omega_1,...,\omega_k$ on the
image.
So the deformation will be defined over the ring $\tilde {{\bf
K}}={\bf
C}[[q_1,...,q_k]]$.

We denote the moduli space of the vertical maps  of degree $d$
  by ${\cal M}_d^{(v)}$ . One can see that ${\cal M}_d^{(v)}$
projects onto $B$ and the fiber is ${{\cal M}(F)}_d$. Let
us denote this projection by $\tau$.

Now we can define vertical correlation functions in the same way we
did before
to define \qucf.
We choose $s$ generic points $x_1,...,x_s$ on ${\bf CP}^1$.
This gives us a map from ${\cal M}_d^{(v)}$ to $X^s$.
Denote it by $\mu$. Let $p_1,p_2,...,p_s$ be
closed differential forms on $X$ of degree $r_1,...,r_s$
respectively. Denote by $P$ the differential form
 ${\pi_1}^*{p_1}\wedge{\pi_2}^*{p_2}\wedge...\wedge{\pi_s}^*{p_s}$.

\vskip 2mm
{\bf Definition.}
{\it
The vertical correlation function
${\langle}p_1|p_2|...|p_s{\rangle}_{(v)}$
is given by
\eqn\vcf{
{\langle}p_1|p_2|...|p_s{\rangle}_{(v)}=\sum_{d}^{} q^d
{\tau}_{!}({\mu}^*P)
}
where ${\tau}_!$ means the integration over fibers. It takes values
in the
cohomology of $B$, i.e. in $H^*(B,{\tilde {{\bf K}}})$.
}

\vskip 2mm
In the same way as for the definition \qucf of the ordinary
correlation
functions, one can prove correctness of this definition.

\vskip 2mm
{\bf Properties of the vertical correlation functions:}

\item{1.} They are multi-linear and skew-symmetric.
\item{2.}
% THE FOLLOWING LINE CANNOT BE BROKEN BEFORE 80 CHAR
${\langle}p_1|p_2|...|p_s|1{\rangle}_{(v)}=
{\langle}p_1|p_2|...|p_s{\rangle}_{(v)} $
 (where 1 is considered as a zero degree differential form).
\item{3.}
% THE FOLLOWING LINE CANNOT BE BROKEN BEFORE 80 CHAR
${\langle}p_1|p_2|...|p_s{\rangle}_{(v)}|_{q=0}=
{\pi}_!(p_1{\wedge}...{\wedge}p_s)$.
\item{4.} ${\langle}p{\rangle}_{(v)}={\pi}_! (p).$
\item{5.} ${\langle}p_1|p_2{\rangle}_{(v)}={\pi}_! (p_1{\wedge}p_2).$
\item{6.} Let $p$ be any differential form on $B$. Then
% THE FOLLOWING LINE CANNOT BE BROKEN BEFORE 80 CHAR
${\langle}{\pi}^*p{\wedge}p_1|p_2|...|p_s{\rangle}_{(v)}=
p{\wedge}{\langle}p_1|p_2|...|p_s{\rangle}_{(v)}$
\vskip 2mm

All properties except the last one can be proved in the same way as
in the
ordinary case.
The last property is quite obvious and follows from the functoriality
and
the properties of the functors ${\tau}_!$, ${\pi}_!$ and ${\pi}^*$.

We can define the vertical quantum cohomology now. The Poincar\'e
pairing on
$X$ gives us an isomorphism of $H^*(X,{\bf C})\otimes{\tilde {{\bf
K}}}$ with
its
dual as $\tilde {{\bf K}}$-modules. In order to define the product we
must have
a triple pairing with values in $\tilde {{\bf K}}$. We already have a
triple
pairing \vcf
with values in $H^*(B,{\tilde {{\bf K}}})$. So we can define the
auxiliary
triple pairing by means of  integration of \vcf over the fundamental
cycle $B$:
\eqn\trip{
{\langle}p_1|p_2|p_3{\rangle}_{auxiliary}=\int _{B} ^{}
{{\langle}p_1|p_2|p_3{\rangle}_{(v)}}.
}
This gives us a product. Certainly this product is skew-commutative.
It is not obvious at all that this product is associative,
but we will prove it below. Let us denote this product by $*$
and the algebra of vertical quantum cohomology by $QH^*_V(X,\tilde
{{\bf K}})$
 (or sometimes by $QH^*_V(X;B,\tilde {{\bf K}})$, in order to avoid
confusion).

First of all, let us notice the following property of
 the multiplication $*$ which shows that it is compatible
with  the vertical correlation functions \vcf:
$${\langle}p_1|p_2|p_3{\rangle}_{(v)}={\langle}p_1*p_2|p_3{\rangle}_{(
v)}.$$
This follows directly from the definition.

The proof of the associativity is based on the relation between
the quantum cohomology of $X$ and the vertical quantum cohomology.
One can get the following formula from the definitions:
% THE FOLLOWING LINE CANNOT BE BROKEN BEFORE 80 CHAR
$${\langle}p_1|p_2|...|p_s{\rangle}|_{q_{k+1}=0,...,q_l=0}=
{\langle}p_1|p_2|...|p_s{\rangle}_{(v)}.$$
This shows that the vertical quantum product is equal to the usual
quantum
product on $X$
when we set  $q_{k+1}=0,...,q_l=0$.  Since the usual quantum product
is
associative,
 this ends the proof of associativity. This relation
between the vertical quantum cohomology and the usual quantum
cohomology
is a particular case of a certain property of the vertical quantum
cohomology,
the so-called  $induction$,
which we will discuss a bit  later.

\subsec{{ Properties of the vertical quantum cohomology.}}

{\bf 1. Product}

Suppose that we have two locally trivial algebraic fiber bundles
$(X_1,B_1,F_1)$ and $(X_2,B_2,F_2)$  for which the
vertical quantum cohomology is defined.
The product $(X,B,F)=(X_1{\times}X_2,B_1{\times}B_2,F_1{\times}F_2)$
of these
bundles is
also a locally trivial algebraic fiber bundle.
Then for the vertical quantum cohomology of the product\foot {One can
see that
the vertical quantum cohomology
of X is a module over
${\tilde {{\bf K}}}_1{\otimes_{\bf C}}{\tilde {{\bf K}}}_2$.} we have
a formula
\eqn\qupro{
QH^*_V(X,{\tilde {{\bf K}}}_1{\otimes_{\bf C}}{\tilde {{\bf K}}}_2)=
QH^*_V(X_1,{\tilde {{\bf K}}}_1){\otimes_{\bf C}}QH^*_V(X_2,{\tilde
{{\bf
K}}}_2).
}
This statement is trivial; it follows immediately from the fact
that the moduli spaces of vertical maps of ${\bf CP}^1$ to $X$
are the products of the moduli spaces for $X_1$ and $X_2$ (~and that
the
Poincar\'e
pairing on the cohomology of $X$ is the product of the Poincar\'e
pairings
on $X_1$ and $X_2$~).

{\bf 2. Restriction}

Consider a locally trivial algebraic bundle $(X,B,F)$ and an
algebraic
map from $\bar{B}$ to $B$. Take the Cartesian product of $X$ and
$\bar{B}$ over $B$. Denote this product by $Y$ and the natural map
from $Y$ to
$X$ by $\nu$. Then $Y$ is a locally
trivial algebraic fiber bundle over $\bar{B}$ with the fiber $F$.
The cohomology $H^*(Y,{\bf C})$ is generated by the
cohomology of $\bar{B}$ and the image of $H^*(X,{\bf C})$ (~this is
true
since the spectral sequence for $Y\rightarrow \bar {B}$ degenerates
in $E_2$ (Deligne's theorem see \Del,\GH)~).
We have the part $\omega_1,...,\omega_k$ of a basis in
$H^2(X,{\bf Z})\cap{H^{1,1}(X,{\bf C})}$. We can choose
${\bar {\omega}}_1={\nu}^*(\omega_1),...,{\bar
{\omega}}_k={\nu}^*(\omega_k)$
as the corresponding part of a basis in $H^2(Y,{\bf
Z})\cap{H^{1,1}(Y,{\bf
C})}$.
Then we get an isomorphism of the rings ${\tilde {\bar {{\bf
K}}}}={\tilde
{{\bf K}}}$.
Now we can say that we have a map:
$$QH^*_V(X,{\tilde {\bf K}}){\rightarrow}QH^*_V(Y,{\tilde {\bar {{\bf
K}}}}).$$
This map is just ${\nu}^*$ (since we have a natural isomorphism of
${\tilde
{\bf K}}$-modules
$QH^*_V(X,{\tilde {\bf K}})$ and $H^*(X,{\tilde {\bf K}})$ ).
Moreover, if we know the vertical cohomology of $X$, then we can
compute
the vertical cohomology of $Y$.

Certainly, we have a map between the vertical quantum cohomology as
${\tilde
{\bf K}}$-modules.
We must show that it respects the multiplication. But this is easy to
see,
since
${\bar {{\cal M}}}_d^{(v)}$ is equal to the  Cartesian product of
${\cal
M}_d^{(v)}$ and
${\bar {B}}$ over $B$. So we get:
% THE FOLLOWING LINE CANNOT BE BROKEN BEFORE 80 CHAR
$${\langle}{\nu}^*p_1|{\nu}^*p_2|...|{\nu}^*p_s{\rangle}_{(v)}=
{\nu}^*{\langle}p_1|p_2|...|p_s{\rangle}_{(v)}.$$
Now we will use the fact that the vertical quantum product
is uniquely defined by
${\langle}p_1*p_2|p_3{\rangle}_{(v)}={\langle}p_1|p_2|p_3{\rangle}_{(v
)}$
for any $p_3$. It is easy to check that
${\langle}{\nu}^*(p_1*p_2)|{\bar
{p}}{\rangle}_{(v)}={\langle}{\nu}^*p_1|{\nu}^*p_2|{\bar
{p}}{\rangle}_{(v)}$
for any ${\bar{p}}$  (here we use the fact that the usual cohomology
is
generated by the cohomology of ${\bar {B}}$ and
the image of $H^*(X,{\tilde {\bf K}})$ ) .

{\bf 3. Induction.}

Suppose that we have  locally trivial algebraic bundles $(X,Y,F_1)$
 and  $(Y,B,F_2)$.
Assume that the algebraic bundle $X{\rightarrow}B$
with the fiber $F$ is locally trivial.
 We  choose a basis in $H^2(X,{\bf Z})\cap{H^{1,1}(X,{\bf C})}$
in the following way: first, we choose a basis
${\tilde{\omega}}_1,...,{\tilde{\omega}}_m$ for $F_1$;
second, we choose a basis
${\bar {\omega}}_{1+m},...,{\bar {\omega}}_{k+m},{\bar
{\omega}}_{k+m+1},...,{\bar {\omega}}_{l+m}$
for $Y$ as before
( when we were defining the vertical correlation functions ).
If we denote now by $i$ an inclusion of $F_1$ into $X$ and by $\pi$
the projection from $X$ onto $Y$, then we can choose a basis
$\omega_1,...\omega_{l+m}$ in
$H^2(X,{\bf Z})\cap{H^{1,1}(X,{\bf C})}$ such that
$i^*(\omega_j)={\tilde{\omega}}_j$ for $1{\leq}j{\leq}m$ and
$\omega_j=\pi^*({\bar {\omega}}_j)$ for $j>m$. Then denote by
${\bf K}$ the ring ${\bf C}[[q_1,...,q_{k+m}]]$ and by
$\tilde {{\bf K}}$ the ring ${\bf C}[[q_1,...,q_m]]$.

In this situation the following relation between the rings
$QH^*_V(X;B,{\bf
K})$ and $QH^*_V(X;Y,{\tilde {{\bf K}}})$ holds:
$$QH^*_V(X;B,{\bf K}){\Big/}\Bigl(
(q_{m+1},...,q_{m+k})QH^*_V(X;B,{\bf
K})\Bigr)=QH^*_V(X;Y,{\tilde {{\bf K}}})$$
$$where ~(q_{m+1},...,q_{m+k})~ is~an~ideal~in~{\bf K}~generated
{}~by~q_{m+1},...,q_{m+k}.$$

This fact follows from two simple observations.
First (~we have already used this many times~), the product $*$ is
defined
uniquely by the correlation functions
${\langle}p_1*p_2|p_3{\rangle}_{(v)}={\langle}p_1|p_2|p_3{\rangle}_{(v
)}$.
Second, the moduli space ${{\cal
M}(X,B)}_{(d_1,...,d_m,0,...,0)}^{(v)}$
coincides with the moduli space ${{\cal
M}(X,Y)}_{(d_1,...,d_m)}^{(v)}$.
So we see that
% THE FOLLOWING LINE CANNOT BE BROKEN BEFORE 80 CHAR
$${\langle}p_1|p_2|...|p_s{\rangle}_{(v)}(X,B)|_{q_{m+1}=0,...,q_{m+k}
=0}=
{\langle}p_1|p_2|...|p_s{\rangle}_{(v)}(X,Y).$$

This complets the sketch of the proof of induction.

\vskip 2mm

{\bf Remark.}
If we look at the definition \vcf of the vertical correlation
functions, we will see that one can define them in the following
situation. Let $G$ be a group acting on an algebraic variety $F$,
$P$ a principal $G$ bundle over $B$ (we are not assuming here that
$P$ is algebraic), and $P{\times_G}F$  our
fiber bundle. Then we can say that the moduli space of the vertical
maps of
degree $d$ is $P{\times_G}{{{\cal M}(F)}_d}$. (~To make this
statement precise
we need to show that all the integrals converge; nevertheless it is
useful to
have this picture in mind.~)
We will define equivariant quantum cohomology in the next section.
One can think about them as a vertical quantum cohomology
where $P$ is $EG$ and $B$ is $BG$ of the Lie group $G$.
This picture and the properties of the vertical quantum
cohomology explain the properties of the equivariant
quantum cohomology.

\newsec {Equivariant quantum cohomology.}

 Equivariant quantum cohomology were defined first in the
paper \Gi. Our approach is based on another, certainly equivalent,
definition.

 We use the setup of the previous chapter;
$X$ is our algebraic variety. Suppose that we
have an  algebraic group $G$ acting on $F$.
Denote by $ g $ the Lie algebra of this group.
As before, we denote by ${\cal M}_d$ the moduli
space of degree $d$ maps from  ${\bf CP}^1$
to $F$. Then the group $G$ acts on ${\cal M}_d$.

Recall that the usual equivariant cohomology is defined as
the cohomology of the following complex:
$${C^k=(\oplus_{i+2j=k}^{} {S^j({ g })^*{\otimes}{\Omega}^i(F)})}^G$$
with the differential:
$$ (d_{ g }(\alpha))(h)=d(\alpha(h))-i(h)(\alpha (h))~~~h{\in}{ g
}.$$
The integration over the
 fundamental cycle of $F$ gives a map $\int$ from the equivariant
cohomology of
$X$ to the equivariant
cohomology of the point ($H^*_G(pt.,{\bf C})$).
This map sends $C^k$ to $(S^{k-{  { dim}}F}({ g })^*)^G$.
We denote by the symbol ${\int_{{\cal M}_d}^{}}$ the map from the
equivariant
cohomology of ${\cal M}_d$ to the equivariant cohomology of the
point.

This shows how we should define the equivariant correlation
functions.
Let us choose $s$ generic points $x_1,...,x_s$ on ${\bf CP}^1$.
This gives us a map $\psi$ from ${\cal M}_d$ to $F^s$.
For $s$ equivariant closed differential forms $p_1,p_2,...,p_s$
(i.e. elements of our complex $C^*$)  we denote by $P$ a product
$P={\pi_1}^*{p_1}\wedge{\pi_2}^*{p_2}\wedge...\wedge{\pi_s}^*{p_s}.$

\vskip 2mm

{\bf Definition.}
 {\it
The equivariant correlation function
${\langle}p_1|p_2|...|p_s{\rangle}_G$
is defined by
\eqn\eqqucf{
{\langle}p_1|p_2|...|p_s{\rangle}_G=\sum_{d}^{} q^d {\int_{{\cal
M}_d}^{}}
{\psi}^*(P).
}
}

By the same argument as in the case
of the usual quantum cohomology, one can see that this definition is
correct.
The properties of the equivariant
correlation functions are similar to those of the vertical
correlation
functions. Nevertheless we formulate them below.

\vskip 2mm
{\bf Properties of the equivariant correlation functions:}

\item{1.} They are multi-linear and skew-symmetric.
\item{2.}
${\langle}p_1|p_2|...|p_s|1{\rangle}_G=
{\langle}p_1|p_2|...|p_s{\rangle}_G $
 ( where 1 is considered as a zero degree equivariant differential
form ).
\item{3.} ${\langle}p_1|p_2|...|p_s{\rangle}_G|_{q=0}={\int}
(p_1{\wedge}...{\wedge}p_s)$ .
\item{4.} ${\langle}p{\rangle}_G={\int } (p).$
\item{5.} ${\langle}p_1|p_2{\rangle}_{(v)}={\int } (p_1{\wedge}p_2).$
\item{6.} Let $p$ be any element of $(S^*({ g })^*)^G$.  If we
multiply $p$ by
1 --- a zero degree G-invariant differential form on $F$ --- we can
consider
it as an element of the complex $C^*$. Then
% THE FOLLOWING LINE CANNOT BE BROKEN BEFORE 80 CHAR
${\langle}p{\wedge}p_1|p_2|...|p_s{\rangle}_G=
p{\langle}p_1|p_2|...|p_s{\rangle}_G$

\vskip 2mm

Let us explain a relation between the vertical correlation functions
and the
equivariant
correlation functions. Suppose that  we have an algebraic principal
$G$-bundle
$P$ over $B$.
Assume that $P$ and $B$ are so nice that the vertical correlation
functions
are defined for $X=P{\times_G}F$ (that is, $X$ is a locally trivial
algebraic
bundle with
the fiber $F$ over $B$).
Let us choose a connection on  $P$ represented
 by  a 1-form $\omega {\in}{({\Omega(P,{ g })})^G}$.
In this situation we have the following
 homomorphisms of differential graded algebras (see theorem 7.42 in
\ref\Getz{Nicole Berline, Ezra Getzler,
                  Michele Vergne, {\it Heat kernels and Dirac
operators},
Berlin; New York: Springer-Verlag, 1992, 369 p.})
$$ {\phi}_{\omega}(X):(C^*,d_{ g })\rightarrow({\Omega}(X),d)$$
$$ {\phi}_{\omega}({\cal M}_d):(({S^*({ g
})^*{\otimes}{\Omega}^*({\cal
M}_d)})^G,d_{ g })\rightarrow({\Omega}({\cal M}_d^{(v)}),d)$$
$$ {\phi}_{\omega}:((S^*({ g })^*)^G,0){\rightarrow}({\Omega}(B),d)$$
These homomorphisms respect everything in the sense that
if we take $s$ equivariant closed differential forms
$p_1,p_2,...,p_s$ on $F$,
then the relation we are talking about is
\eqn\reveeq{
{\phi}_{\omega}
% THE FOLLOWING LINE CANNOT BE BROKEN BEFORE 80 CHAR
\bigl({\langle}p_1|p_2|...|p_s{\rangle}_G\bigr)=
{\langle}{\phi}_{\omega}(X)(p_1)|{\phi}_{\omega}(X)(p_2)|...|{\phi}_{\
omega}(X)(p_s){\rangle}_{(v)}.
}
One can check this property using functoriality of the homomorphisms
 ${\phi}_{\omega}(X)$, ${\phi}_{\omega}({\cal M}_d)$,
${\phi}_{\omega}$ and
others
(see paragraph 7.6 in \Getz).

Now we can  define equivariant quantum cohomology.
We have a spectral sequence $E_2^{p,q}=H^p(BG,H^q(F,{\bf K})$
converging to the
equivariant
cohomology of $F$. Note that it degenerates in the term $E_2$ ( see
\Del).
So $H_G^*(F,{\bf K})$ is a free module over $H^*(BG,{\bf K})=(S^*({ g
})^*)^G{\otimes_{\bf C}}{\bf K}$.
The usual pairing ${\int } (p_1{\wedge}p_2)$ with values in
$H^*(BG,{\bf K})$
is non-degenerate
(since it respects the spectral sequence in natural way see
\ref\Ginzburg{V.~Ginzburg, {\it Equivariant cohomology and K\"ahler
geometry}
Funct.~Anal.~Appl. {\bf 21:4} (1987), 271-283.}).
 We want to define a
multiplication satisfying the following property:
$$ {\langle}p_1*p_2|p_3{\rangle}_G={\langle}p_1|p_2|p_3{\rangle}_G.
$$
 Since ${\langle}p_1|p_2{\rangle}_G={\int } (p_1{\wedge}p_2)$ we see
that
if the quantum multiplication exists, then it is unique.
We can think of  the triple $G$--equivariant pairing as a morphism
of  $H^*(BG,{\bf K})$-modules:
\eqn\trieq{
H_G^*(F,{\bf K}){\otimes_{H^*(BG,{\bf K})}}H_G^*(F,{\bf
K}){\otimes_{H^*(BG,{\bf K})}}H_G^*(F,{\bf
K}){\rightarrow}H^*(BG,{\bf K}).
}
Together with the non-degenerate Poincar\'e pairing, the triple
pairing \trieq
defines  a product
\eqn\equpro{
H_G^*(F,{\bf K}){\otimes_{H^*(BG,{\bf K})}}H_G^*(F,{\bf
K}){\rightarrow}H_G^*(F,{\bf K}).
}
We denote by $QH_G^*(F,{\bf K})$ the algebra defined by this product.

One can check that this definition is correct and that the product
\equpro is
associative.
 For example, associativity of the product can be proved by comparing
\equpro
with
the vertical quantum product (~using \reveeq~). One only needs to
notice that
there exists the a map from $QH_G^*(F,{\bf K})$ to
$QH^*_V(P{\times_G}F,{\bf
K})$ which
respects the multiplication.

So we have the equivariant quantum cohomology ring $QH_G^*(F,{\bf
K})$.
Moreover,
$QH_G^*(F,{\bf K})$ is an algebra over $H^*(BG,{\bf K})$.
The following commutative diagram describes the relation between the
equivariant and the vertical quantum cohomology:

\eqn\diag{
\matrix{ &QH_G^*(F,{\bf K}) &{\longrightarrow}
&QH^*_V(P{\times_G}F,{\bf K})
\cr
&{\Big \uparrow} &  &{\Big \uparrow} \cr
&H^*(BG,{\bf K}) &{\longrightarrow} &H^*(B,{\bf K}) \cr
}}

Now we discuss some properties of the equivariant quantum cohomology
(see \Gi).

\vskip 2mm

{\bf Product.}
Let $F'$ and $F''$ be $G'$- and $G''$-spaces respectively
($G'$, $G''$ are affine algebraic groups, etc. ).
We have the  cohomology groups  $QH_{G'}^*(F',{\bf K}')$ and
$QH_{G''}^*(F'',{\bf K}'')$.
Now  take the product $F'{\times}F''$ and consider it as a
$G'{\times}G''$-space.
Consider  the ring $QH_{G'{\times}G''}^*(F'{\times}F'',{\bf
K}'{\otimes_{\bf
C}}{\bf K}'')$.
We can ask what is the relation between the equivariant quantum
cohomology of
$F'$,  $F''$
and that of $F'{\times}F''$.
The answer is given by the formula:
\eqn\qecprod{
QH_{G'{\times}G''}^*(F'{\times}F'',{\bf K}'{\otimes_{\bf C}}{\bf
K}'')=QH_{G'}^*(F',{\bf K}'){\otimes_{\bf C}}QH_{G''}^*(F'',{\bf
K}'').
}

This formula follows immediately from the the definitions.

\vskip 2mm

{\bf Restriction.}
Let $F$ be a $G$-space and suppose that we have an affine algebraic
subgroup
$\tilde {G}$ of $G$.
Note that the Leray spectral sequence
$E_2^{p,q}=H^p_G\bigl(pt,H^q(F,{\bf K})\bigr)$
(~${\tilde {E}}_2^{p,q}=H^p_{\tilde {G}}(pt,H^q(F,{\bf K})$~)
degenerates in the term $E_2$ (~respectively ${\tilde {E}}_2$~).
We have a natural morphism
of the rings $f:~H^*_G(F,{\bf K}){\rightarrow}H^*_{\tilde {G}}(F,{\bf
K})$
which comes from a map
$$ ({S^*({ g })^*{\otimes}{\Omega}^*(F)})^G{\rightarrow}({S^*({\tilde
{ g
}})^*{\otimes}{\Omega}^*(F)})^{\tilde {G}}$$
 induced by the inclusion ${\tilde { g }}\hookrightarrow{ g }$.
Let us make the following remark: $H^*_{\tilde G}(F,{\bf K})$ is
generated
by $H^*_G(F,{\bf K})$ as a $H^*_{\tilde {G}}(pt,{\bf K})$-module.
Denote the natural map from $H^*_G(pt,{\bf K})$ to $H^*_{\tilde
{G}}(pt,{\bf
K})$ by $\delta$.
If we look now at the definition of the equivariant correlation
functions, we
can easily see from functoriality (~paragraph 7.6 of \Getz~) that
$${\langle}f(p_1)|f(p_2)|...|f(p_s){\rangle}_{\tilde {G}}=\delta
({\langle}p_1|p_2|...|p_s{\rangle}_G).$$
 Since the equivariant quantum product is uniquely defined by the
correlation
functions and because of the fact that $H^*_{\tilde G}(F,{\bf K})$ is
generated
by $H^*_G(F,{\bf K})$ as a $H^*_{\tilde {G}}(pt,{\bf K})$-module, the
map $f$ respects the equivariant quantum product. So we have a
morphism
of rings:
$$f:~QH^*_G(F,{\bf K}){\rightarrow}QH^*_{\tilde {G}}(F,{\bf K}).$$

Using the fact that the spectral sequence ${\tilde {E}}_*$
degenerates in
${\tilde {E}}_2$,
we see that $QH^*_{\tilde {G}}(F,{\bf K})$ is completely defined by
$QH^*_G(F,{\bf K})$, $H^*_{\tilde {G}}(F,{\bf K})$ and the morphism
$f$. If
$QH^*_G(F,{\bf K})$ is equal to a quotient of
an $H^*_G(pt,{\bf K})$--algebra $A$  modulo
an ideal $I$, then $QH^*_{\tilde {G}}(F,{\bf K})$ should be
equal to the quotient of  $A{\otimes_{\bf K}}H^*_{\tilde {G}}(pt,{\bf
K})$
modulo the image of $I$.

\vskip 2mm

Now we are going to discuss two particular cases of restriction.
\item{1.} Let $\tilde {G}$ be the trivial group. Then
$$QH^*_{\tilde {G}}(F,{\bf K})=QH^*(F,{\bf K}).$$
So we see that the quantum cohomology of $F$ is
equal to the quotient of the equivariant
quantum cohomology modulo an ideal generated by
the equivariant cohomology of the point. That is
\eqn\qecp{
QH^*(F,{\bf K})=QH^*_G(F,{\bf K})/(H^*_G(pt,{\bf K}))
}
\item{2.} Let $H$ be any affine algebraic group,
$K$ its maximal compact subgroup, and $G$ the corresponding
reductive algebraic subgroup. Let $F$ be a $H$-space.
Then $F$ naturally becomes a $G$-space.
 We claim that the
following isomorphism takes place:
\eqn\qecsg{
QH^*_{G}(F,{\bf K})=QH^*_{H}(F,{\bf K})
}
Really, we have $H^*_{H}(F,{\bf K})=H^*_G(F,{\bf K})$
(~since $F$ is a projective variety~) and
$H^*_{H}(pt,*)=H^*_G(pt,*)$.
{}From the restriction property we get
$$QH^*_{\tilde {G}}(F,{\bf K})=QH^*_G(F,{\bf K})$$
as $H^*_{H}(pt,*)=H^*_G(pt,*)$--algebras.
\itemitem{2.1.} Let $G$ be a parabolic subgroup of some
semisimple group and $\tilde {G}$ be its
Levi subgroup. There is a natural
projection from $G$ onto $\tilde {G}$.
Let $F$ be a $\tilde {G}$-space. Then
$F$ naturally becomes a $G$-space.
So we get (~see \qecsg~)
\eqn\qecps{
QH^*_{G}(F,{\bf K})=QH^*_{\tilde {G}}(F,{\bf K})
}

\vskip 2mm

{\bf Induction.}
Let ${\tilde {G}}$ be an affine algebraic subgroup of $G$ with a
simply-connected compact K\"ahler quotient $G/{\tilde {G}}$. Let
$F$ be a ${\tilde {G}}$-space and a simply-connected
compact K\"ahler variety $X=G{\times_{\tilde {G}}}F$.
One can wonder about the relations
between $QH^*_{\tilde {G}}(F,\tilde {{\bf K}})$ and $QH^*_G(X,{\bf
K})$.
The answer is as  follows.

First, the space $X$ is fibered over  $G/{\tilde {G}}$ with a fiber
$F$.
We have a basis ${\tilde{\omega}}_1,...,{\tilde{\omega}}_k$
 in $H^2(F,{\bf Z})\cap{H^{1,1}(F,{\bf C})}$. We can choose a
basis $\omega_1,...,\omega_l$ as we did in Section 3 when we
described
the vertical quantum cohomology.
There exists a natural map from
$((S^*({ g })^*{\otimes}{\Omega}(X))^G,d_{ g })$ to
$((S^*(\tilde { g })^*{\otimes}{\Omega}(F))^{\tilde {G}},d_{\tilde {
g }})$
which induces an isomorphism of the equivariant cohomology.

Let us describe this map.
We can treat $(S^*({g})^*{\otimes}{\Omega}(X))^G$ as
${\rm Hom}_G(S^*({g}),{\Omega}(X))$.  Indeed, if given a map $T$ then
for any
$v{\in}S^*({g})$ we have $T(v){\in }{\Omega}(X)$. There is a
canonical
inclusion of $F$ into $X$ as a fiber over $1{\in}(G/{\tilde {G}})$.
So we can restrict the section $T(v)$ to $F$:
$$T(v)|_F{\in}{\Omega}(X)|_F={\Lambda}^*({g}/{\tilde {g}})^*
{\otimes_{\bf C}}{\Omega}(F).$$
To an element $T{\in}{\rm Hom}_G(S^*({g}),{\Omega}(X))$ we can
associate an element $$\tilde {T}{\in}{\rm Hom}_{\tilde
{G}}(S^*({g}),
{\Lambda}^*({g}/{\tilde {g}})^*{\otimes_{\bf C}}{\Omega}(F)).$$
Thus we get a map from $(S^*({g})^*{\otimes}{\Omega}(X))^G$
to $(S^*({g})^*{\otimes_{\bf C}}{\Lambda}^*({g})^*/{\tilde
{g}}){\otimes_{\bf
C}}{\Omega}(F))^{\tilde {G}}$. There is a natural map from
$S^*({g})^*{\otimes_{\bf C}}{\Lambda}^*({g}/{\tilde {g}})^*$
to  $S^*(\tilde {g})^*$  induced by inclusion of
$S^*(\tilde {g})$ into $S^*({g}){\otimes_{\bf
C}}{\Lambda}^*({g}/{\tilde
{g}})$. The composition of these two maps gives us the desired
one. One can check that it respects the differential.

Using these natural maps, one can see that
$${\langle}p_1|p_2|...|p_s{\rangle}_{\tilde
{G}}={\langle}p_1|p_2|...|p_s{\rangle}_G|_{q_{k+1}=0,...,q_l=0}.$$
Finally, there is a natural isomorphism $H^*_{\tilde
{G}}(F,*)=H^*_G(X,*)$, we
obtain
\eqn\qecind{
QH^*_{\tilde {G}}(F,{\tilde {{\bf K}}})=QH^*_G(X,{\bf
K})|_{q_{k+1}=0,...,q_l=0}
}
This is an isomorphism of the rings, but there are additional
structures. The
left hand side is an
$H^*_{\tilde {G}}(pt,{\tilde {{\bf K}}})$-algebra and the right hand
side
is an $H^*_G(pt,{\tilde {{\bf K}}})$-algebra (since ${\tilde {{\bf
K}}}=
{\bf K}/(q_{k+1},...,q_l)$). Let us describe the relation between
these
structures. One can see that there is a natural morphisms of the
rings
$H^*_G(pt,{\tilde {{\bf K}}})\rightarrow H^*_{\tilde {G}}(pt,{\tilde
{{\bf
K}}})$. Thus the left hand side is also an $H^*_G(pt,{\tilde {{\bf
K}}})$-algebra.
 One can see that isomorphism \qecind is the isomorphism of
$H^*_G(pt,{\tilde {{\bf K}}})$-algebras.
relation
We want to make a remark that this property
of the equivariant quantum cohomology is an analog
of the induction for the vertical quantum cohomology.
One can prove this property
reducing it to the same property of the vertical quantum
cohomology.

\vskip 3mm

{\bf Remark.}
The equivariant quantum cohomology gives us
a simple way to compute the vertical quantum cohomology.
If we had a principal $G$-bundle $P$ and a $G$-space $F$
then  in order to  compute the vertical quantum cohomology
of $P{\times_G}F$ it would be sufficient to know the equivariant
quantum
cohomology of $F$.

\newsec{Computation of the equivariant quantum cohomology of partial
flag
manifolds.}
\subsec{Classical cohomology ring.}
Let us consider a partial flag manifold $F_{n_1\cdots n_k}$. There
are $k$
canonical vector bundles $\xi  _1,\ldots ,\xi  _k$ on this variety.
To describe
them we introduce $k$ auxiliary vector bundles $\eta _1,\ldots ,\eta
_k$. A
fiber of $\eta _i$ over a flag $\{0\subset V^1\subset \cdots \subset
V^k\}\in
F_{n_1\cdots n_k}$ is a vector space $V^i$. In particular, $\eta _k$
is a
trivial bundle. The canonical bundles $\{\xi  _i\}$ are defined by
\eqn\canbun{
\xi _i=\cases{\eta _i/\eta _{i-1}, &if $i>1$; \cr
              \eta _1, &if $i=1$. \cr
              }
}
We denote by $x^{(i)}_j$ the $j$-th Chern class of $\xi  _i$.

The cohomology ring of the flag manifold is a quotient of the
polynomial ring
${\bf C}[x_1^{(1)},\ldots ,x_{n_k}^{(k)}]$ modulo the ideal generated
by the
coefficients of the polynomial
\eqn\polyn{
P(\lambda )= \lambda ^{n} -\prod_{i=1}^k(\lambda ^{n_i} + \lambda
^{n_i-1}x_1^{(i)} +\cdots + x_{n_i}^{(i)})
}
We choose the basis $\omega _1=c_1(\eta _1),\ldots ,\omega
_{k-1}=c_1(\eta
_{k-1})$ in $H^2(X, {\bf Z})\cap H^{1,1}(X, {\bf C})$.

Our partial flag manifold is a quotient $GL(n, {\bf C})/P$ where $P$
is a
corresponding parabolic subgroup.  Let $\alpha _1,\ldots ,\alpha _n$
be the
simple positive roots of $sl(n, {\bf C})$. To each $\alpha _i$ we can
canonically associate a subgroup $SL_{(i)}(2, {\bf C})\subset GL(n,
{\bf C})$.
If $i\neq n_1$, $i\neq n_1+n_2$, $\ldots$, $i\neq n_{1}+\ldots
+n_{k-1}$ then
$SL_{(i)}(2, {\bf C})$ lies in the parabolic subgroup $P$. On the
other hand,
if $i=n_1+\ldots +n_{j-1}$ then
$SL_{(i)}(2, {\bf C})$ intersects $P$ along the maximal torus ${\bf
C}^*\subset
SL_{(i)}(2, {\bf C})$. So we get $k-1$ canonical embeddings of ${\bf
CP}_{(i)}^1=SL_{(i)}(2, {\bf C})/{\bf C}^*$ into $F_{n_1\cdots n_k}$.

The homological type of curve $\Sigma \subset F_{n_1\cdots n_k}$ is
given by
the string $d=(\omega _1(\Sigma ),\ldots ,\omega _{k-1}(\Sigma ))$ of
its
coordinates. In particular, the homological type of ${\bf
CP}_{(n_1+\cdots
+n_{i})}^1$ corresponding to the simple root $\alpha _{n_1+\cdots
+n_{i}}$ is
\eqn\strco{
(\overbrace{0,\ldots,0,}^{i-1}\, 1, \overbrace{0,\ldots, 0}^{n-i})
}

The group $GL(n, {\bf C})$ acts on the manifold $F_{n_1\cdots n_k}$.
Let us
describe the {\it equivariant} cohomology $H^*_{GL(n)}(F_{n_1\cdots
n_k},{\bf
K})$.
As an algebra over $H^*_{GL(n)}(pt,{\bf K})={\bf K}[c_1,\ldots ,c_n]$
it is
generated by the elements $\{y^{(i)}_j\}$. The map
$H^*_{GL(n)}(F_{n_1\cdots
n_k},{\bf K})\rightarrow H^*(F_{n_1\cdots n_k},{\bf K})$ sends
$\{y^{(i)}_j\}\longrightarrow \{x^{(i)}_j\}$. The relations are given
now by
the coefficients of the polynomial
\eqn\polequi{
P_{GL(n)}(\lambda )= \lambda ^{n} + c_1 \lambda ^{n-1} +\ldots + c_n
-\prod_{i=1}^k(\lambda ^{n_i} + \lambda ^{n_i-1}x_1^{(i)} +\cdots +
x_{n_i}^{(i)})
}

\subsec{The equivariant quantum cohomology of Grassmanians}
The equivariant quantum cohomology ring\foot{For the Grassmanian the
ring of
coefficients is generated by a single parameter $q$: ${\bf K}={\bf
C}[[q]]$.}
$QH_{GL(n)}^*(Gr(n,m),{\bf K})$ of Grassmanian
$Gr(n,m)=U(n)/(U(n-m)\times
U(m))$ as an algebra over $H^*_{GL(n)}(pt,{\bf K})$ is generated by
$y^{(1)},\ldots,y^{(1)}_m,y^{(2)}_1,\ldots, y^{(2)}_{n-m}$. By Lemma
B2 (see
Appendix B), the relations in $QH_{GL(n)}(Gr(n,m))$ are {\it the
deformations}
of the classical relations \polequi. In order to compute them we need
to know
the quantum cohomology ring $QH^*(Gr(n,m),{\bf K})$. It was discussed
in many
papers \gep -- \WiG. In the last reference it is shown that this ring
is
generated by $x^{(1)},\ldots,x^{(1)}_m,x^{(2)}_1,\ldots,
x^{(2)}_{n-m}$ and the
 relations are the coefficients of the following polynomial equation:
\eqn\polgr{
(\lambda ^m + \lambda ^{m-1}x^{(1)}_1 +\cdots +x^{(1)}_m)(\lambda
^{n-m} +
\lambda ^{n-m-1}x^{(2)}_1 +\cdots +x^{(2)}_{n-m})=\lambda ^n +
(-1)^{n-m}q
}
{}From Lemma B1 (see Appendix B) or \ref\SP{S.~Piunikhin,
unpublished.} it
follows that the {\it whole} ideal of relations in $QH^*(Gr(n,m),{\bf
K})$ is
generated by \polgr.

Now we can find the relations in $QH_{GL(n)}^*(Gr(n,m))$ we are
after. Note
that the degree of $q$ is equal to $2n$. Therefore only the highest
($2n$)
degree relation in \polequi is deformed:
\eqn\hyprel{
y_m^{(1)}y_{n-m}^{(2)}=c_n + \alpha q
}
where $\alpha$ is some constant. This constant is determined using
the
restriction property (see section 4). One can see it equals the
corresponding
constant  $(-1)^{n-m}$ in \polgr.
Note that this answer is consistent with the general formula \reeqc
for the
equivariant quantum cohomology of the partial flag manifolds.

\subsec{The key lemma}
According to Lemma B2
$$QH^*_{GL(n)}(F_{n_1\cdots n_k},{\bf K})={\bf K}[y_1^{(1)},\ldots
,y_{n_k}^{(k)}; c_1,\ldots ,c_n]/QI_{GL(n)}$$
where the ideal $QI_{GL(n)}$ is generated by some deformations of the
classical
relations \polequi.
The following Lemma is a trivial generalization of the Lemma 4.2.1
\Gi.

{\bf Lemma} {\it For $k>2$, suppose that a quasi-homogeneous relation
of the
form
$$\lambda ^{n} +  c_1 \lambda ^{n-1} +\ldots + c_n
-\prod_{i=1}^k(\lambda
^{n_i} + \lambda ^{n_i-1}x_1^{(i)} +\cdots +
x_{n_i}^{(i)})=O(q_1,\ldots
,q_{k-1})[\lambda ,q, y, c ]$$
is satisfied in the equivariant quantum cohomology algebra of the
partial flag
manifold  $F_{n_1\cdots n_k}$ modulo $q_i$ for each $i=1,\ldots,
k-1$. Then the
relation holds identically (i.~e.~ for all $q$).}

{\it Sketch of the proof.} Indeed, since the $LHS$ of the relation in
question
is homogeneous of degree $2n$ (we choose $deg\; \lambda$=2), the
hypothesis of
Lemma 1 means that the difference $LHS-RHS$ is divisible by
$q_1\ldots
q_{k-1}$. But $deg\; q_i=n_{i}+n_{i+1}$ and
$$deg\; q_1\ldots q_{k-1}=2n-n_1-n_k>n\ {\rm for\;} k>2.$$

Q.E.D.

\subsec{Relations in $QH^*_{GL(n)}(F_{n_1\cdots n_k},{\bf K})$}
We are going to prove that the relations \reeqc generate the ideal
$QI_{GL(n)}$. We will do this by induction on $k$.
\item{1.)} (A base of induction.) When $k=2$, the partial flag
manifold
$F_{n_1\ n_2}$ is called the Grassmanian $Gr(n_1+n_2,n_2)$. We have
already
proven \reeqc in that case.
\item{2.)} (A step of induction.) Suppose that our assumption is true
for all
$k<k_0$. Let us prove it for $k=k_0$. According to the key lemma, it
suffices
to prove these relations modulo $q_i$ for all $i=1,\ldots k-1$.

Let us notice that\foot{A symbol $F_n$ which may appear in this
formula stands
for a point $\{pt\}$.}
\eqn\flpres{
F_{n_1\cdots n_k}=GL(n, {\bf C})\times _{P_j}(F_{n_1\cdots n_j}\times
F_{n_{j+1}\cdots n_k})
}
where $P_j\subset GL(n, {\bf C})$ is a corresponding parabolic
subgroup.
Using the induction property \qecind   we get
\eqn\qcfl{
QH^*_{GL(n, {\bf C})}(F_{n_1\cdots n_k},{\bf
K})|_{q_j=0}=QH^*_{P_j}(F_{n_1\cdots n_j}\times F_{n_{j+1}\cdots
n_k},\tilde{\bf K})
}
where $\tilde{\bf K}={\bf K}/((q_j))$.
{}From the restriction property \qecps   it follows that
\eqn\qcinduc{
QH^*_{P_j}(F_{n_1\cdots n_j}\times F_{n_{j+1}\cdots n_k},\tilde{\bf
K})=
QH^*_{GL(N_1, {\bf C})\times GL(N_2, {\bf C})}(F_{n_1\cdots
n_j}\times
F_{n_{j+1}\cdots n_k},\tilde{\bf K})
}
where $N_1=n_1+\ldots +n_j$ and $N_2=n_{j+1}+\ldots +n_k$.
Finally, from the product \qecprod we have
\eqn\qccc{\eqalign{
QH^*_{GL(N_1, {\bf C})\times GL(N_2, {\bf C})}&(F_{n_1\cdots
n_j}\times
F_{n_{j+1}\cdots n_k},\tilde{\bf K})= \cr
&=QH^*_{GL(N_1, {\bf C})}(F_{n_1\cdots n_j},{\bf K}_1)
\otimes _{\bf C}
QH^*_{GL( N_2, {\bf C})}(F_{n_{j+1}\cdots n_k},{\bf K}_2) \cr
}}
and  $\tilde{\bf K}={\bf K}_1\otimes {\bf K}_2$.
Taking \qcfl -- \qccc   together we obtain the natural isomorphism
\eqn\qcfinisom{
QH^*_{GL(n, {\bf C})}(F_{n_1\cdots n_k},{\bf K})|_{q_j=0}=
QH^*_{GL(N_1, {\bf C})}(F_{n_1\cdots n_j},{\bf K}_1)
\otimes _{\bf C}
QH^*_{GL(N_2, {\bf C})}(F_{n_{j+1}\cdots n_k},{\bf K}_2).
}
We want to stress that \qcfinisom is {\it an isomorphism of
$QH^*_{GL(n, {\bf
C})}(pt,\tilde{\bf K})$--algebras}.
This isomorphism together with the assumption 2.) shows that  for
$k=k_0$  the
relations \reeqc   hold modulo $q_j$. Since it is true for all
$j=1,\ldots,
k-1$, the step of induction is complete.

Q.E.D.

{\bf Remark.} Using the restriction property \qecp we find that the
{\it
quantum cohomology} ring $QH^*(F_{n_1\cdots n_k},{\bf K})$ is
generated by
$x_1^{(1)},\ldots ,x_{n_k}^{(k)}$ of degrees ${\rm deg}\
x_i^{(j)}=2i$ with the
relations given by the coefficients of the polynomial
\eqn\ree{
{\rm det}[\lambda \,{\bf 1}+A]-\lambda ^{n}
}
where the $n\times n$ matrix $A$ is defined  in the Section 2 just
below the
formula \reeqc.

\newsec{Appendix A}
\subsec{Remarks on the quantum cohomology $QH^*({\bf P}(E))$.}

Whereas the {\it vertical} quantum cohomology of $F_{n_1\cdots
n_k}(E)$ are
described by \rerel, we do not now if it is possible to compute {\it
the
quantum cohomology} ring of the flag bundle in general. It seems to
be quite a
nontrivial problem, as two examples below show.

{\bf Example 1}. Take a projective space ${\bf P}^m$ as a base, a
vector bundle
$E=\oplus_{i=0}^{r_i}{\cal O}(r_i)$, $r_0\geq \cdots \geq r_n$ on it
and let
$X={\bf P}(E)$. A K\"ahler cone of ${\bf P}(E)$ is two-dimensional
and is
generated by its two edges $y\in H^{11}({\bf P}^m) \subset
H^{11}({\bf P}(E))$
and $z=x+r_0y$ where $x\in H^{11}({\bf P}(E))$ is a generator as in
\cpvb and
$r_0={\rm max}\{r_i\}$. The cohomology ring $H^*({\bf P}(E))$ is
generated by
$x$ and $z$ with the relations
\eqn\clrelI{\eqalign{
&\, y^{m+1}=0 \cr
&\prod_{i=0}^n[z+(r_i-r_0)y]=0
}}
The first Chern class $c_1({\bf P}(E))=nz + (m-\sum (r_0-r_i))y$ is
positive
iff
$k=m-\sum (r_0-r_i)>0$. The Novikov ring ${\bf K}$ is generated by
two elements
(quantum deformation parameters) $q_v$ ("vertical") and $q_h$
("horizontal") of
degrees $2(n+1)$ and $2k$ respectively.

The relations in the quantum cohomology ring $QH^*({\bf P}(E))$ in
this example
can be obtained using a fact $P(E)$ is a toric variety \Ba.
Equivalently, one
finds the relations from the symplectic reduction of a linear
$\sigma$-model on
the $n+m+2$-dimensional vector space by the action of $U(1)\times
U(1)$ \WiNII.
 The quantum relations are:
\eqn\qlrelI{\eqalign{
&\, y^{m+1}=q_h\prod_{i=0}^n [z+(r_i-r_0)y]^{r_0-r_i} \cr
&\prod_{i=0}^n[z+(r_i-r_0)y]=q_v
}}

{\bf Example 2}. The quantum cohomology of the flag manifold
$F_3=U(3)/(U(1)\times U(1)\times U(1))$ are computed in \ref\Sa{
V.Sadov, {\it
On equivalence of Floer's and quantum cohomology.}
Preprint HUTP-93/A027.},\Gi. They are generated by two elements $x$
and $y$ of
degree two. The edges of the two-dimensional K\"ahler cone of $F_3$
are $y$ and $x+y$.  The quantum
relations are:
\eqn\qfltr{\eqalign{
&x^2+xy+y^2=q_v+q_h \cr
&y^3=q_h(2y+x) \cr
}}
A space $F_3$ can be realized as a projectivization ${\bf P}(E)$ of a
rank 2
irreducible vector bundle on ${\bf P}^2$, then $y$ is a generator of
the ring
$H^*({\bf P}^2)$ and $x$ is a generator\foot{The Chern classes  of
$E$ are
$c_1(E)=y$ and $c_2(E)=y^2$} as in \cpvb.
The first Chern class of $F_3$ is $c_1(F_3)=2(x+2y)$ so the degrees
of both
$q_v$ and $q_h$ are 4.

In both examples we see, that $QH^*({\bf P}(E))$ is not a
$QH^*(B)$-algebra.
For $q_h\neq 0$ the relations for $y$ in \qlrelI, \qfltr differ from
the
relation in the quantum
cohomology of projective space. This is a general situation in
quantum
cohomology; only in the classical limit "on the base" ($q_h=0$ in our
examples)
does the ``universality", like in \cpvb, get restored and the ring
$QH^*({\bf
P}(E))$ becomes  an algebra over $QH^*(B)|_{class}=H^*(B)$. But then
it just
coincides with the vertical quantum cohomology $QH_V^*({\bf
P}(E),{\bf P}^n,B)$
as defined above.

It can happen sometime, like it happens in \qlrelI, that the
"vertical"
relation for $x$ is the same in both rings $QH_V^*({\bf P}(E),{\bf
P}^n,B)$ and
 $QH^*({\bf P}(E))$. But the second example shows that in general
this is not
the case and in $QH^*({\bf P}(E))$ the "vertical" relations
$QH_V^*({\bf
P}(E),{\bf P}^n,B)$ get deformed by $q_h$ as well the "horizontal"
relation do.

\newsec{ Appendix B.}
We use the following setup.
We have a compact K\"ahler algebraic variety $X$; $c$ is the first
Chern class of the tangent bundle to $X$. We denote by $q^d$ an
element of the group ring of the sublattice $L$
in $H_2(X,{\bf Z})\cap H_2(X,{\bf C})$
 generated by the homology classes of rational
complex curves
(${\bf CP}^1$).
 Let us choose some basis $\omega_1,...,\omega_k$ in $H^2(X,{\bf
Z}){\cap}
H^{1,1}(X,{\bf C})$ such that the homology class of a curve $S$ is
given by a
string
$d=(d_1,...,d_k)$ of its coordinates with respect to
$\omega_1,...,\omega_k$
with $d_1,...,d_k{\geq}0$.
Then the element $q^d$ can be identified with the monomial
${q_1}^{d_1}...{q_k}^{d_k}$.

Let us define a ring of coefficients ${\bf K}={\bf
C}[[q_1,q_2,...,q_k]]$. This
is a local ring; let $I$ be its maximal ideal.
The quantum cohomology is an algebra over the ring ${\bf K}$.
Moreover, by
definition it is a free
module over ${\bf K}$. The ring ${\bf K}$ is graded so that
deg$\,q^d=c(d)$ (here we consider $d$ as an element of $H_2(X,{\bf
C})$).

Let $p_1,...,p_l$ be some cocycles of degree $r_1,...,r_l$
respectively.
Then a correlation function ${\langle}p_1|p_2|...|p_l{\rangle}$ with
values in ${{\bf K}}$ is defined. We assume that
${\langle}p_1|p_2|...|p_l{\rangle}$ is always
homogeneous of degree $r_1+r_2+...+r_l-{ {\rm dim}}(X)$.
Notice that the pairing ${\langle}p_1|p_2{\rangle}$
is ordinary Poincar\'e pairing.

Let $x_1,...,x_r$ be  homogeneous generators of the ring $H^*(X,{\bf
C})$. Let
$f_1,...,f_s$ be generators of the ideal of relations  in $H^*(X,{\bf
C})$ that
is,
$f_1,...,f_s$ are homogeneous polynomials in  $x_1,...,x_r$ and
$H^*(X,{\bf C})=S\Lambda[x_1,...x_r]/(f_1,...,f_s)$.
Then the following lemma holds.

{\bf Lemma 1. }
{\it
There exist  polynomials $g_i(x_1,...,x_r,q_1,...q_k)~~for~~
i=1,...,s$
such that $g_i(x_1,...x_r,0,...0)=0~~i=1,...,s$  with the following
relations  in $QH^*(X,{\bf K})$
$$f_i(x_1,...,x_r)=g_i(x_1,...,x_r,q_1,...q_k).$$
In this situation the quantum cohomology ring equals
$$QH^*(X,{\bf K})=S\Lambda_{\bf
K}[x_1,...x_r]/(f_1-g_1,...,f_s-g_s).$$
}

{\bf Proof:}
Let us choose some homogeneous basis $y_1,...,y_u$ in the vector
space
$H^*(X,{\bf C})$.
Let $z_1,...,z_u$ be the dual basis with respect to Poincar\'e
duality.
Let $f_i(x_1,...,x_r)=\sum_{j_1,...,j_r}
{a_{{j_1},...,{j_r}}x_1^{j_1}...x_r^{j_r}}$ and
let $N$ be such positive integer that $ a_{{j_1},...,{j_r}}=0$ for
$j_1+...+j_r>N$.
Using the fact that \foot{We can avoid using this fact since the
quantum
multiplication is a deformation of the ordinary one.
We know that
$x_1^{j_1}*...*x_r^{j_r}=x_1^{j_1}...x_r^{j_r}~~(~mod~~I~)$. So,
instead of writing correlation functions,
we can notice that
% THE FOLLOWING LINE CANNOT BE BROKEN BEFORE 80 CHAR
${\langle}x_1^{j_1}*...*x_r^{j_r}|y_j{\rangle}=
{\langle}x_1^{j_1}...x_r^{j_r}|y_j{\rangle}~~
(~mod~~I~)$. As follows from the proof, knowing this we can avoid the
formula
${\langle}p_1|p_2|...|p_l{\rangle}=
{\langle}p_1*p_2*...*p_{l-1}|p_l{\rangle}$.}
${\langle}p_1|p_2|...|p_l{\rangle}=
{\langle}p_1*p_2*...*p_{l-1}|p_l{\rangle}$
we can write
$$\sum_{j_1,...,j_r}
% THE FOLLOWING LINE CANNOT BE BROKEN BEFORE 80 CHAR
{a_{{j_1},...,{j_r}}{\langle}\underbrace{x_1|x_1|...|x_1}_{j_1}|
\underbrace{x_2|...|x_2}_{j_2}|...|\underbrace{x_r|...|x_r}_{j_r}
|\underbrace{1|1|...|1}_{N-(j_1+...+j_r)}|y_j{\rangle}}= $$
$$=(f_i(x_1,...,x_r),y_j)+h_{i,j}(q_1,...,q_k)=h_{i,j}(q_1,...,q_k),$$

$$~~where~~h_{i,j}(q_1,...,q_k)~~is~~polynomial~~in~~q_1,...,q_k~~~and
 ~~
h_{i,j}(0,...,0)=0.$$

So we see that in $QH^*(X,{\bf K})$
$$f_i(x_1,...,x_r)=\sum_{j=1}^u {h_{i,j}(q_1,...,q_k)z_j}.$$

Certainly we have a map from $S\Lambda[x_1,...x_r]$ to $QH^*(X,{\bf
K})$ (as
${\bf K}$-modules).
Moreover it is surjective since this map respects grading, the map
from
$S\Lambda[x_1,...x_r]/(I)$ to  $QH^*(X,{\bf K})/(I)$
is surjective and we can use  Nakayama's lemma. So we can express
 $z_j{\in}QH^*(X,{\bf K})$ as a polynomial of $x_1,...,x_r$ with
coefficients
in ${\bf K}$.
This proves the first part of the lemma.

We have a map from $S\Lambda[x_1,...x_r]/(f_1-g_1,...,f_s-g_s)$ to
$QH^*(X,{\bf
K})$.
These two ${\bf K}$-modules both are finitely generated and the
isomorphism
$$S\Lambda[x_1,...x_r]/(f_1-g_1,...,f_s-g_s)/(I)=QH^*(X,{\bf
K})/(I)$$ holds.
It follows now from the Nakayama's lemma that these modules are
isomorphic:
$$S\Lambda[x_1,...x_r]/(f_1-g_1,...,f_s-g_s)=QH^*(X,{\bf K}).$$

Q.E.D.

This lemma shows that if we know the deformed relations $f_i=g_i$
 then the quantum cohomology ring is
$QH^*(X,{\bf K})=S\Lambda_{\bf K}[x_1,...x_r]/(f_1-g_1,...,f_s-g_s)$.

Let us state the similar lemma for the equivariant quantum
cohomology.
We will not prove it since the proof is a simple adaptation of the
proof
of Lemma 1.

 Let G be affine algebraic group acting on $X$. Let
$y_1,...,y_r$ be homogeneous generators of the algebra
$H^*_G(X,{\bf C})$ over the ring $H^*_G(pt,{\bf C})$.
Let $c_1,...,c_n$ be some homogeneous generators of the ring
 $H^*_G(pt,{\bf C})$.
Let ${\bar {f}}_1,...,{\bar {f}}_s$ be generators of the ideal of
 relations between $y_i$. That is, ${\bar {f}}_1,...,{\bar {f}}_s$
are
homogeneous polynomials in $y_1,...,y_r$ and $c_1,...,c_n$ and
$$H^*_G(X,{\bf C})=S{\Lambda}_{H^*_G(pt,{\bf C})}[y_1,...,y_r]/({\bar
{f}}_1,...,{\bar {f}}_s).$$
\smallskip
{\bf Lemma 2.}
{\it
   There exist polynomials ${\bar
{g}}_i(y_1,...,y_r,c_1,...,c_n,q_1,...,q_k)$
for $i=1,...,s$ such that ${\bar
{g}}_i(y_1,...,y_r,c_1,...,c_n,0,...,0)=0$
for $i=1,...,s$ and the following relations in $QH^*_G(X,{\bf K})$
hold
$${\bar {f}}_i(y_1,...,y_r,c_1,...,c_n)={\bar
{g}}_i(y_1,...,y_r,c_1,...,c_n,q_1,...,q_k).$$
In this case the equivariant quantum cohomology ring
$$QH^*_G(X,{\bf K})=S{\Lambda}_{H^*_G(pt,{\bf K})}[y_1,...,y_r]/
v({\bar {f}}_1-{\bar {g}}_1,...,{\bar {f}}_s-{\bar {g}}_s)$$
as $H^*_G(pt,{\bf K})$ algebra.
}

\newsec{ Appendix C.}

  We keep here the setup of appendix {\bf B}.
Let us notice that the ring ${\bf K}$ ( the ring of ``coefficients'')
depends
on the
choice of the basis in $H^2(X,{\bf Z}){\cap}
H^{1,1}(X,{\bf C})$. The reason is that the natural ring of
``coefficients'' $R$
is embedded into the ring ${\bf K}$ (the ring ${\bf K}$ is bigger
then the
natural
one). We use the ring ${\bf K}$ to simplify the explanations. Now we
are going
to describe the ring $R$.

Let $H$ be a semigroup in $H_2(X,{\bf Z})\cap H_2(X,{\bf C})$
generated by the homology classes of the rational complex curves
(${\bf
CP}^1$).
Denote by $\tilde {R}={\bf C}[H]$ the semigroup ring of $H$. There is
a natural
homomorphism from
$\tilde {R}$ to $\bf C$ which sends $0{\in}H$ to $1$ and all other
elements of $H$ to zero. We denote the kernel of this map by $J$;
$J$ is a maximal ideal.
The ring $R$ is a completion of the ring $\tilde {R}$ with respect to
the ideal $J$:
$$R=\hat {\tilde {R}}.$$
The quantum cohomology of $X$ ($QH^*(X,R)$) is an algebra over $R$.
One can see that there is a natural embedding of the ring $R$
 into the ${\bf K}$ and that
 $QH^*(X,{\bf K})=QH^*(X,R){\otimes_R}{\bf K}$. All the statements
in this paper are true over the ring $R$ although they are formulated
over the
ring $\bf K$.

Moreover, if $c$ is positive then all our statements are true over
the ring
$\tilde {R}$ but we are not going to discuss that.

\newsec{ Appendix D.}

We are going to show that the moduli spaces of algebraic maps of
${\bf CP}^1$ (~with three marked points~) to the projective variety
$X$ exist.
\item{1.} For $X={\bf CP}^N$ this fact is obvious.
\item{2.} This statement is also trivial for $X={\bf P}^{N_1}\times
\ldots
\times {\bf P}^{N_s}$ since the moduli space of the product is the
product of
the moduli spaces.
\item{3.} For general $X$ let us choose an integral basis $\omega
_1,\ldots,\omega _s$ of $H^{1,1}(X)$ such that $\omega _i$ belongs to
the
K\"ahler cone of $X$ for $i=1,\ldots, s$. Let ${\cal
L}_1,\ldots,{\cal L}_i$ be
the corresponding invertible sheaves. There exists $M\in {\bf N}$
such that all
the sheaves ${\cal L}_1^M,\ldots,{\cal L}_i^M$ are very ample. Let us
take the
corresponding embeddigs $\sigma _1,\ldots, \sigma _s$ of $X$ into
${\bf
P}^{N_1},\ldots,{\bf P}^{N_s}$. Then the embedding $\Sigma =\sigma
_1\times
\ldots \times \sigma _s:X\hookrightarrow Z={\bf P}^{N_1}\times \ldots
\times
{\bf P}^{N_s}$ has the following property: for any two algebraic
curves $S_1$
and $S_2$ in $X$ such that $[S_1]\neq[S_2]$ (~$[S_1],\,[S_2]\in
H_2(X,{\bf
C})$~) the pushforwards $\Sigma _*[S_1]\neq \Sigma _*[S_2]$ (~$
\Sigma
_*[S_1],\,\Sigma _*[S_2]\in H_2(Z,{\bf C})$~). Let $d\in H_2(X,{\bf
C})\cap
H_2(X,{\bf Z})$. Denote by ${\cal M}_{\Sigma _*{d}}(Z)$ the moduili
space of
the agebraic maps ${\bf CP}^1\rightarrow Z.$
An algebraic map ${\bf CP}^1\rightarrow X$ gives a point in ${\cal
M}_{\Sigma
_*{d}}(Z)$. Thus we can identify the space of algebraic maps ${\bf
CP}^1\rightarrow Z$ as a subspace in ${\cal M}_{\Sigma _*{d}}(Z)$.
One can see
that the condition that the map $\phi :{\bf CP}^1\rightarrow Z$
factors through
a map $\tilde \phi :{\bf CP}^1\rightarrow X$ so that $\phi =\Sigma
\circ \tilde
\phi$ is an algebraic condition. So the space ${\cal M}_{\Sigma
_*{d}}(X)$ of
algebraic maps ${\bf CP}^1\rightarrow X$ is an algebraic subvariety
in ${\cal
M}_{\Sigma _*{d}}(Z)$.

\listrefs
\bye